\documentclass[a4paper,12pt]{article}
\usepackage{epsfig} 
\usepackage{boxedminipage}
\usepackage{float}
\usepackage{cite}
\usepackage{array}

\newcommand {\grtapprox}{\ 
  \raisebox{-1ex}{$\stackrel{\textstyle>}{\sim}$}\ }
\newcommand{\lessapprox}{\ 
  \raisebox{-1ex}{$\stackrel{\textstyle<}{\sim}$}\ }

\begin{document}

\begin{titlepage}
\title{Osmotic Stabilisation of Concentrated Emulsions and Foams}
\author{A.J. Webster, M.E. Cates$^*$\\
Department of Physics and Astronomy, University of Edinburgh,\\
JCMB, King's Buildings, Mayfield Road, Edinburgh,\\ 
EH9 3JZ, U.K.}
\end{titlepage}
\maketitle

\begin{abstract}
\normalsize
In the absence of coalescence, coarsening of emulsions (and foams) is controlled by molecular diffusion of the dispersed phase species from one emulsion droplet (or foam bubble) to another. 
Previous studies of dilute emulsions have shown how the osmotic pressure
of a trapped species within droplets
can overcome the Laplace pressure differences that drive coarsening, and
``osmotically stabilise'' the emulsion. 
Webster and Cates ({\sl Langmuir}, {\bf 1998}, {\sl 14},
$2068-2079$) gave rigorous criteria
for osmotic stabilisation of mono- and polydisperse emulsions, in the dilute regime.
We consider here whether analogous criteria exist for 
the osmotic stabilisation of mono- and polydisperse 
{\sl concentrated} emulsions and foams.
We argue that in such systems the pressure differences driving coarsening are small compared to the mean Laplace pressure.
This is confirmed for a monodisperse 2D model, for which an exact
calculation gives the pressure in bubble $i$ as $P_i = P + \Pi + P_i^G$,
with $P$ the atmospheric and $\Pi$ the osmotic pressure, and 
$P_i^G$ a `geometric pressure' that reduces to the Laplace pressure only
for a spherical bubble, and depends much less strongly on bubble deformation than the Laplace pressure itself. In fact, for Princen's 2D emulsion model,
$P_i^G$ is only 5\% larger in the dry limit than the dilute limit.
We conclude that osmotic stabilisation of dense systems typically requires a
pressure of trapped molecules in each droplet that is comparable to the Laplace pressures the same droplets would have if they were spherical, as opposed to the 
much larger Laplace pressures actually present in the system.
We study the coarsening of foams and dense emulsions when there is insufficient of the trapped species present. Various rate-limiting mechanisms are considered, and their domain of applicability
and associated droplet growth rates discussed. In a concentrated foam or emulsion,
a finite yield threshold for droplet rearrangement among stable droplets may be enough to prevent coarsening of the remainder.

\end{abstract}


\section{Introduction}

Previous
work\cite{Higuchi,Hallworth,Davis,Davis2,Kab87a,Kab87b,Reiss,Kuehmann,Webster} 
on dilute emulsions has shown 
that, when coarsening is solely caused by a diffusive flux of dissolved 
dispersed-phase molecules between droplets (Ostwald ripening\cite{Ostwald}),
coarsening may be prevented by the addition of a sufficient number of
molecules that are insoluble in the continuous phase and hence
{\sl trapped} within droplets. 
The trapped molecules provide an
osmotic pressure which counteracts the Laplace pressure due to surface
tension\cite{Webster} (which drives coarsening); resulting in
`osmotic stabilisation'. 
A quantitative criterion for stability, valid
even for
emulsions with polydispersity in both the droplets' sizes and
number of trapped molecules they contain, was given in \cite{Webster}. 
It was also found in \cite{Webster} that for an 
`insufficiently stabilised' emulsion (without enough trapped species to obey
the required criterion), the subsequent coarsening 
was qualitatively unaltered from that without any of the trapped
species \cite{Lifshitz,Wagner}. The only effect was is to reduce
the effective volume fraction of the coarsening droplets by an amount
corresponding to the final volume of a population of small droplets
that attain coexistence with the coarsened bulk phase.
The latter are prevented from entirely dissolving by the trapped molecules they
contain. When the effective volume fraction is reduced to zero, full stability
is achieved.

In the present paper we discuss whether analogous conditions 
exist for the osmotic stabilisation of concentrated emulsions and foams containing trapped insoluble molecules in the dispersed phase; 
we again consider both the monodisperse and the polydisperse case. 
Previously the use of Nitrogen as a trapped species to stabilise foams
has been investigated with theory\cite{Falls,Gandolfo},
experiment\cite{Falls,Parr,Bamforth,Gandolfo}, and computer 
simulations\cite{Weaire_Pageron}, but theoretical conditions for
stability of foams with nonspherical bubbles were not found. The use of other
trapped gases to inhibit dissolution (as opposed to coarsening) of spherical bubbles is addressed in\cite{ultrasound} where the effect of condensation
of the included gases is also addressed; we do not consider this here.
In our work, we treat the idealised case of a fully insoluble trapped species (perfect trapping) whereas in many cases the results will be modified by
residual solubility effects (considered for the dilute case in\cite{Webster}).
Some gases are, however, practically insoluble in water, for example C$_2$F$_6$
\cite{Stone}; and in the case of emulsions, the insoluble limit is easily
achieved (by using oligomeric species). For simplicity we do not consider
the effects of residual solubility in the present work.

From now on we use mainly the language of foams, in which the trapped and soluble gas are treated as ideal, but the work is also applicable
to dense emulsions whose droplets contain an ideal mixture of soluble
and trapped molecules. The only distinction between these cases is
that foams are compressible, but in fact much of the paper concerns
{\sl effectively incompressible} foams, in
which pressure differences between bubbles are negligible compared
with their mean pressures, and the total bubble volume is conserved. 

The osmotic stabilisation of foams is a far more complex proposition
than for dilute emulsions. 
In a dilute emulsion, each droplet is spherical with an internal
pressure that is increased by the droplets Laplace pressure, which
is directly related to the bubble's volume through its spherical
geometry, and equals $2\sigma/R_i$ for a droplet of radius $R_i$ and
surface tension $\sigma$. 
Since nonspherical foam bubbles in contact 
have no simple relationship between their
volume and surface area, 
there is no direct relationship, in a concentrated foam, 
between a bubble's pressure and its volume.   

A bubble $i$ with a radius of curvature $r_i$ at its Plateau borders has 
a Laplace pressure of $2\sigma/r_i$, and an internal pressure of
$P_i=P+2\sigma/r_i$. Given a connected domain of the liquid phase, the
atmospheric pressure $P$ equals the pressure in the Plateau
borders (we ignore the effects of gravity throughout).   
The drier the foam, the smaller the borders and the larger the Laplace
pressure, so that if osmotic stabilisation of a foam or emulsion required the
pressure of trapped molecules to balance the Laplace pressure (as it does in the
dilute case) osmotic stabilisation would be hard to achieve.
However, another relevant length scale is
$R_i \equiv ({3}V_i/{4\pi})^{1/3}$, where $V_i$ is the volume
of bubble $i$. Since the curvature of adjacent bubble-bubble faces
are of order $1/R_i$, the pressure differences between adjacent
bubbles (which are responsible for coarsening), are of order
$\sigma/R_i \ll \sigma/r_i$. 
So if the osmotic stabilisation of a foam merely requires the partial
pressure of trapped molecules to balance $\sigma/R_i$, then osmotic
stabilisation is a reasonable proposition. We shall confirm below that
this is so. The reduction in the driving force for coarsening from the level
suggested by the Laplace pressure has long been recognised (see e.g., \cite{Weaire_Pageron}) but the implications of this for stabilisation by
trapped species has not previously been addressed in detail.

In what follows we first give in Section \ref{dilutefoams} some results for the dilute case (based on \cite{Webster}), and then investigate what
factors determine the pressure within concentrated foam
bubbles. We follow the approach of
Princen\cite{Princen1,Princen2,Princen3}, and study 
the osmotic compression (at atmospheric pressure $P$) of previously
spherical foam bubbles by an osmotic pressure $\Pi$. 
Section \ref{compressed/drained} discusses the disjoining pressures 
between bubbles, the uniformity or otherwise of the osmotic
pressure $\Pi$, and the condition for mechanical equilibrium
with an excess bulk gas phase created by coarsening. The increase in a
bubble's pressure above that of a bulk gas is defined by $P_i^G \equiv 
P_i-P-\Pi$, and we argue that typically $P_i^G\sim \sigma/R_i$. 
This is explicitly confirmed in Section \ref{2Dmodel} for Princen's
monodisperse 2D model. We call $P_i^G$ the {\sl geometric pressure}
and identify it (rather than Laplace pressure) as the driving force for coarsening, in general. A formal condition for osmotic stabilisation is derived in Section
\ref{stabfoamcond}, and examined for various limiting geometries.
We discuss $P_i^G$ for polydisperse foams in Section
\ref{2Dpolysystems}, and consider exceptions to our estimate that
$P_i^G\sim \sigma/R_i$ which might arise under certain conditions
(which we argue to be uncommon). 
We then use the geometric pressure to further investigate the
stability requirements of polydisperse foams. 

The coarsening of insufficiently stabilised foams is studied in
Sections \ref{CoarsenFchapt}-\ref{FYS}, starting with a simple
mean-field model in Section \ref{rapidR}.
By considering dissipation rates for the diffusive flux of gas between
bubbles and for the viscous rearrangement of bubbles, we are able to predict (as a function of various parameters) the rate-limiting mechanism and the 
associated growth rate (Section \ref{DissipRates}).
Section \ref{elasticD} shows that when bubble
rearrangements are sufficiently rare, elastic stresses may arrest coarsening. Section
\ref{FYS} extends this to the case of a finite yield strain beyond which bubble
rearrangements will cause the foam to flow, finding that a foam's initial state then determines
whether coarsening will still occur. 
We conclude in Section \ref{conclusions} with a brief summary and
discussion of our results.

\section{Dilute Foams (Spherical Bubbles)}\label{dilutefoams}

A ``dilute foam'' comprises of spherical gas bubbles floating freely in a
solvent. Following \cite{Webster}, we treat the bubbles as macroscopic
objects and neglect their entropy of translation. 

Firstly we consider the size (and hence composition) at which
spherical bubbles, containing both soluble and trapped gas molecules, 
may coexist with a bulk gas phase (for example, formed by one bubble 
in the foam becoming macroscopic). 
The gas pressure $P_i$ within a spherical bubble labelled $i$ is
increased above the atmospheric pressure by its Laplace pressure
$2\sigma/R_i$:
\begin{equation}\label{dilP_i}
P_i = P + \frac{2\sigma}{R_i} 
\end{equation}

We consider bubbles with $N_i^s$ soluble gas molecules and $N_i^T$
trapped gas molecules, and treat the gases as ideal. 
Then $P_i = P_i^s + P_i^T$,  
where $P_i^s$, $P_i^T$ are the partial pressures of the soluble and
trapped gas molecules respectively, and for the soluble gas molecules
\begin{equation}\label{dilmu1}
\mu_i^s = \mu_b^s + kT \ln\left( \frac{P_i^s}{P} \right) 
\end{equation}
where $\mu_b^s$ is the chemical potential of a bulk gas phase of
soluble molecules at atmospheric pressure $P$.  
Using Eqs.\ref{dilP_i},\ref{dilmu1} and $P_i^s=P_i-P_i^T$, we may
write $\Delta \mu_i \equiv \mu_i^s - \mu_b^s$ as 
\begin{equation}
\Delta \mu_i = kT \ln \left( 1 + \frac{2\sigma/R_i - P_i^T}{P}
  \right) 
\end{equation}
So that when $P_i^T = 2\sigma/R_i$, $\mu_i^s = \mu_b^s$ and bubbles
may coexist with a bulk gas phase (at pressure $P$). 
Using the ideal gas law for the trapped species $P_i^TV_i = N_i^T kT$,
such coexistence requires 
\begin{equation}
\frac{2\sigma}{R_i} = \frac{N_i^TkT}{(4\pi/3)R_i^3} 
\end{equation}
where $V_i=(4\pi/3)R_i^3$. This expression determines a `coexistence volume', $V_i^B$,
at which the Laplace pressure and the partial pressure of
trapped gas balance. Solving for $V_i^B$ we have
\begin{equation}\label{dilVB}
V_i^B = \sqrt{\frac{3}{4\pi}}
\left(\frac{N_i^TkT}{2\sigma}\right)^{3/2} 
\end{equation}
This is identical to the coexistence volume for a dilute
incompressible emulsion droplet, of interfacial tension $\sigma$, containing $N_i^T$ molecules of 
a trapped species \cite{Webster}.

At coexistence with a bulk gas phase, $\Delta \mu_i=0$ and $P_i^s =
P_b^s = P$, so from the ideal gas law  
the number of soluble gas molecules in bubbles is 
\begin{equation}\label{NsB3Ddil}
N_i^{sB} = \frac{P}{kT} V_i^{B} = \frac{P}{kT} \sqrt{\frac{3}{4\pi}} \left
  ( \frac{N_i^TkT}{2\sigma} \right)^{3/2}  
\end{equation} 
Similarly in two dimensions, dilute (circular) bubbles will
coexist with a bulk gas phase only if their areas
$\mathcal{A}_i=\mathcal{A}_i^B$, with 
\begin{equation}
\mathcal{A}_i^B = \frac{1}{\pi} \left( \frac{N_i^TkT}{\sigma} \right)^2 
\end{equation}
They then contain $N_i^{sB}$ soluble gas molecules with 
\begin{equation}\label{NsB2Ddil}
N_i^{sB} = \frac{P}{kT} \left( \frac{1}{\pi} \right)
\left( \frac{N_i^T kT}{\sigma} \right)^2 
\end{equation}

\section{Nondilute Foams}\label{compressed/drained}

Now we consider ``nondilute foams'' in which bubbles press on one
another and are distorted into nonspherical shapes\cite{f1}. 
Following Princen\cite{Princen1,Princen2,Princen3}, we
consider the 
compression of a previously dilute foam under an osmotic pressure $\Pi$. 

In a nondilute foam a typical bubble's interface consists of
gently curved faces which contact adjacent bubbles, and highly curved
regions at the Plateau borders. 
Although bubble faces press on one another with a disjoining pressure,
this typically results in a negligible direct contribution to
the free energy\cite{BuzzaCates}; we may 
assume that the disjoining forces only
{\sl indirectly} affects a bubble's free energy, by 
distorting its shape and increasing
surface area. Put differently, throughout the foam the surface tension
$\sigma$ is taken constant, independent of the volume fraction of 
gas present.  

\begin{figure}[htd]
\begin{center}
\leavevmode
\epsfig{file=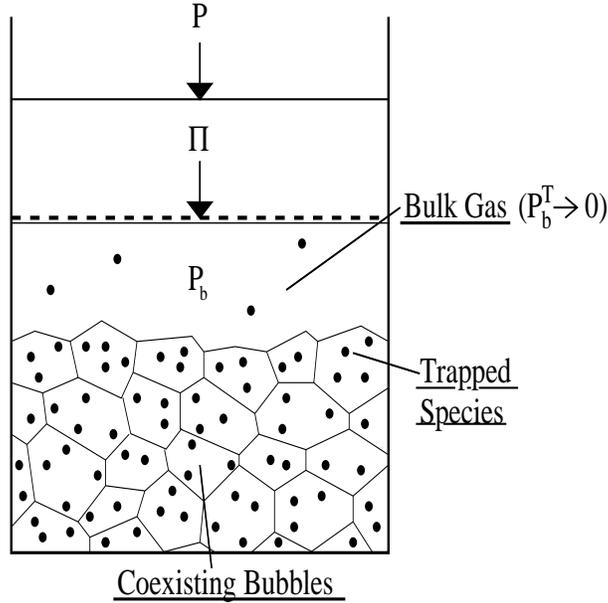,width=8cm,height=8cm}   
\end{center}
\caption{The pressure of a bulk gas $P_b^s$ is negligibly increased by
  interfacial tension, so $P_b^s$ will balance
  (and hence equal) $P+\Pi$. If the bulk gas arises by coarsening, so that one bubble grows to macroscopic size, its own trapped species contribute negligibly to the pressure.}\label{bulkgasP}
\end{figure}

Following Equation \ref{dilmu1} (the ideal gas law) we 
obtain $\Delta \mu_i \equiv
\mu_i^s - \mu_b^s$ for
nondilute bubbles as $\Delta \mu_i = kT \ln \left(P_i^s/P_b^s
\right)$, where $\mu_b^s$ is the chemical potential of a bulk gas of
soluble molecules subject to pressure $P_b^s$.
Were such a bulk gas to arise by coarsening, $P_b^s$ would 
balance both the atmospheric pressure $P$ and 
the osmotic pressure $\Pi$; defining $\tilde{P}\equiv P + \Pi$ we
require $P_b^s=\tilde{P}$ (see figure \ref{bulkgasP}) \cite{f2}. So for
nondilute foam bubbles we have    
\begin{equation}\label{munonideal}
\Delta \mu_i = kT \ln \left( \frac{P_i^s}{\tilde{P}} \right) 
\end{equation}
As we compress the system with a
semipermeable membrane,  the previously spherical
bubbles will distort in shape and the  continuous phase will 
flow\cite{f3}  
so that the additional pressure is evenly distributed amongst
bubbles. (This contrasts with granular materials for
example\cite{Radjai,Fragile}.) 
In the simplest scenario of 2D, monodisperse foams, the
bubbles are compressed into monodisperse hexagons with `rounded'
corners, and equal internal  pressures.  However, compression of a
general  polydisperse foam  will  result in bubbles  deforming into
various  shapes; a given bubble will have a pressure which depends not
only on its volume (as for spherical bubbles), but on the arrangement
and  pressures of {\sl all} of its neighbours. 

The above discussion may be clarified by noting that bubble interfaces
which press on one another, do so with a radius of curvature at most 
of order $1/R_i$ (where $R_i$ is the radius of a bubble with the same
volume in an uncompressed state). Hence pressure differences between
bubbles are of order $\sigma/R_i$. So if $\Pi$ exceeds $\sigma/R_i$ then
the increase in bubbles' pressures will (to within terms of order
$\sigma/R_i$), be  homogeneous throughout the foam \cite{f4}.
So we define the increase in a bubble's pressure above that of bulk
gas ($P_b^s = \tilde P = P + \Pi$), by 
\begin{equation}\label{PGdef} 
P_i^G \equiv P_i - P - \Pi 
\end{equation}
and expect $P_i^G$ to be of order $\sigma/R_i$. Here $P_i^G$ reduces to
the Laplace pressure of the $i$th dilute, uncompressed ($\Pi=0$), spherical 3D
bubble, for which the increase in pressure above a coexisting bulk gas is $P_i^G = P_i-P=2\sigma/R_i$.  
But in a compressed state $P_i^G$ is no longer the Laplace pressure, for the latter is $\sigma/r_i$, with $r_i$ the radius of curvature at a
Plateau border of droplet $i$.

Since we consider an ideal mixture of soluble and trapped gases, 
then Eq. \ref{PGdef} requires $P_i^s = P + \Pi + P_i^G - P_i^T$, which
substituting into Eq. \ref{munonideal} gives 
\begin{equation}
\Delta\mu_i = kT \ln \left( 1 + \frac{P_i^G - P_i^T}{\tilde{P}}
\right) 
\end{equation}
So since a bulk gas formed by coarsening has $\Delta\mu_i=0$, a bubble
can coexist with a coarsened bulk gas phase when $P_i^G=P_i^T$.

We note $P_i^G$ as defined by Eq. \ref{PGdef} is generally not 
easy to calculate from geometric considerations. 
Nonetheless the origin of $P_i^G$ for a spherical bubble is
geometrical, and $P_i^G$ of a nondilute foam bubble is determined by packing
geometry, so we refer to $P_i^G$ as a bubble's ``geometric pressure''. 
To confirm the reasonableness of our arguments for the magnitude of
$P_i^G$, the following section considers a 2D model for which $P_i$,
$\Pi$, and $P_i^G$ are exactly calculable.  

We first clarify what it means for a foam to be ``incompressible''. 
If $P_i^G \ll P +\Pi$ then variations in a bubble's pressure are  
negligible compared with its actual internal pressure. 
So a bubble's gas density is approximately unaffected by its 
geometric pressure, and hence we may treat such bubbles as effectively
incompressible. 
We emphasise that bubbles with $P_i^G\ll P+\Pi$ may only be
treated as incompressible with respect to {\sl coarsening} (which
changes their geometric pressures); such systems are {\sl not}
incompressible under changes in $P$ or $\Pi$ (which changes their
Laplace pressures as well).

\section{A Simple 2D Model}\label{2Dmodel}

\begin{figure}[htd]
\begin{center}
\leavevmode
\epsfig{file=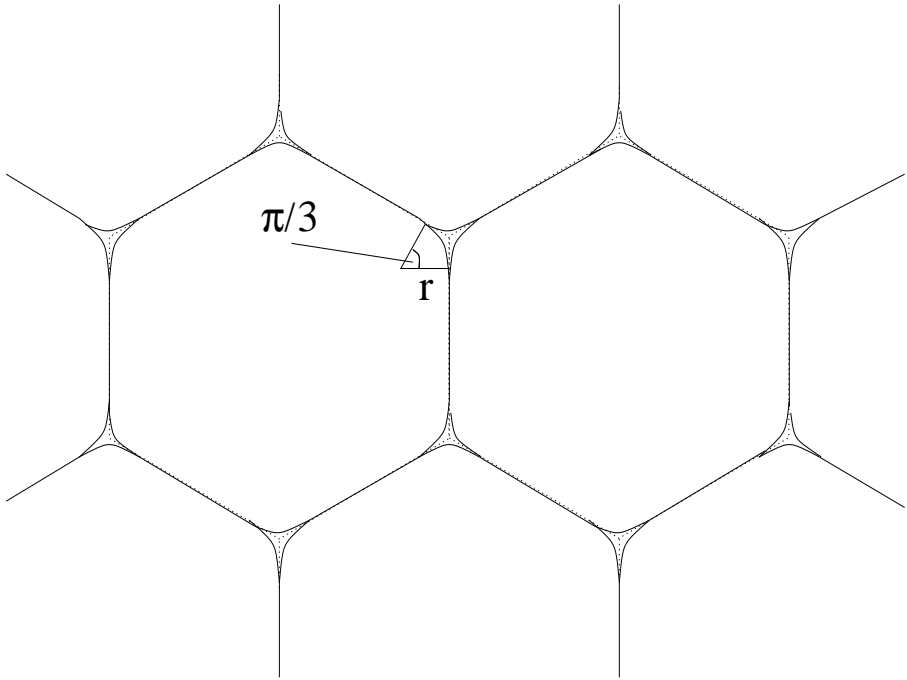,width=12cm, height=9cm}   
\end{center}
\caption{A 2D monodisperse foam. Since bubbles
  are equivalent they have the same pressures and hence radii of
  curvature at their Plateau borders.}\label{2Dmodelfig} 
\end{figure}

We now study a monodisperse, incompressible
2-dimensional foam, 
at osmotic pressure $\Pi$. 
(In Appendix \ref{Fbalance} an alternative argument
confirms the results for an equivalent, but compressible model.) 
For $\Pi \neq 0$ monodisperse bubbles 
will form an hexagonal array, with bubbles distorted into
approximately hexagonal 
shapes but with rounded corners (see figure \ref{2Dmodelfig}). 
For simplicity the film thickness between adjacent bubbles is taken to
be negligible. 
We note that an equivalent system of osmotically compressed,
monodisperse {\sl cylindrical} emulsion droplets was considered by
Princen\cite{Princen1,Princen2,Princen3}. Princen's calculation
for the osmotic pressure is used later. 

Since the bubbles are taken to be incompressible, the 
osmotic pressure $\Pi$ does work by removal of the continuous liquid 
phase from bubbles' Plateau borders. 
Writing the area of liquid associated with each bubble as $\mathcal{A}_{li}$
(with $\mathcal{A}_{li}$ given as the sum of one third of the volume of liquid
at each of its Plateau borders), then since we consider a monodisperse
system with bubbles of area $\mathcal{A}_i$, the osmotic pressure is given
by\cite{WebsterTh,Princen1,Princen2,Princen3}  
\begin{equation}\label{2Dosm}
\Pi = - \sigma \left( \frac{\partial \mathcal{L}_i}{\partial
    \mathcal{A}_{li}} 
\right)_{\mathcal{A}_i} 
\end{equation}
where $\mathcal{L}_i$ is the interfacial length of bubble $i$. 
(We keep a separate label $i$ for each bubble, although they are
identical, for clarity later on.) 

In Eq. \ref{PGdef} we defined $P_i^G \equiv P_i - P - \Pi$.
Since the monodisperse system considered here has hexagonal symmetry,
by analogy with the Laplace pressure of a circular bubble we
propose that $P_i^G$ may be calculated by considering the increase in
a bubble's interfacial length with an isotropic expansion at fixed
liquid area
$\mathcal{A}_{li}$. So for this system we postulate that  
\begin{equation}\label{PGhyp} 
P^G_i = \sigma \left( \frac{\partial \mathcal{L}_i}{\partial
    \mathcal{A}_i} \right)_{\mathcal{A}_{li}}  
\end{equation}
with $(\partial \mathcal{L}_i/\partial \mathcal{A}_i)_{\mathcal{A}_{li}}$
calculated for an isotropic
expansion, and with $P^G_i$ the same for all bubbles (since the system is
monodisperse). 
$P^G_i$ will be different for different lattice arrangements, and also
varies (along with bubble shape), with liquid content. Only in the
absence of bubble-bubble contacts will $P^G_i$ equal the Laplace
pressure, with $P_i^G=\sigma/R_i=\sigma/r_i$.

Since $P^G_i$ is calculated at a fixed
volume of liquid per gas bubble, we have fixed radius
of curvature at the Plateau borders.
Hence Eq. \ref{PGhyp} gives 
\begin{equation}\label{2DGP2}
P^G_i = \sigma \left( \frac{\partial \mathcal{L}_i}{\partial l}
  \right)_r \left 
  ( \frac{\partial l}{\partial \mathcal{A}_i} \right)_r
\end{equation}
where $l$ is the length of the flat bubble-bubble faces (all equal).

In Appendix \ref{bubblegeom} we obtain the exact expressions for the
interfacial  length, and area of a nearly hexagonal bubble as 
\begin{equation}\label{intlength}
\mathcal{L}_i = 6l +2\pi r 
\end{equation}
\begin{equation}\label{areacons}
\mathcal{A}_i= \pi r^2 +6lr + \frac{3\sqrt{3}}{2} l^2 
\end{equation}
After differentiation and some algebra, Eqs \ref{2DGP2},
\ref{intlength}, and \ref{areacons} give  
\begin{equation}\label{2DGP4}
P^G_i = \frac{6\sigma}{\sqrt{36r^2+6\sqrt{3}(\mathcal{A}_i-\pi r^2)}}  
\end{equation}
We note that $P^G_i$ may also be written as 
\begin{equation}\label{PG2} 
P^G_i =
\frac{\sqrt{2\sqrt{3}}\sigma}{\sqrt{\mathcal{A}_i+\mathcal{A}_{li}}}  
\end{equation}
where $\mathcal{A}_{li} = \left( 2\sqrt{3} - \pi \right) r^2$ is the
area of liquid in Plateau borders, per bubble.

Princen\cite{Princen1,Princen2} calculated the osmotic
pressure of monodisperse, cylindrical emulsion droplets by 
equating the work done by the osmotic pressure $\Pi$ with the
increase in interfacial energy as droplets distort (at fixed droplet
volume). For this quasi-2D geometry he
obtained 
\begin{equation}\label{PrincenPi} 
\Pi = \frac{\sigma}{R_i} \left( \frac{\phi}{\phi_0} \right)^{1/2} \left
  [ \left( \frac{1- \phi_0}{1-\phi} \right)^{1/2} - 1 \right] 
\end{equation}
with $\phi$ the volume fraction of emulsion droplets, $\phi_0$ their
volume fraction at first contact,
and $R_i$ the corresponding radius.  
Princen's\cite{Princen1,Princen2} calculation 
applies equally to our {\sl incompressible}, monodisperse 2D foam: the area $A_i$ and volume $V_i$ of monodisperse cylindrical drops of
length $\Lambda$ obey
$A_i=\mathcal{L}_i\Lambda$, $V_i=\mathcal{A}_i\Lambda$. Therefore  
$\phi_0=(\pi R_i^2/2\sqrt{3}R_i^2)$ and
$\phi=\mathcal{A}_i/(\mathcal{A}_i+\mathcal{A}_{li})$.
In the variables used in
this paper we have 
\begin{equation}\label{PrincenVar1}
\left( \frac{\phi}{\phi_0} \right)^{1/2} = \left
  ( \frac{2\sqrt{3}}{\pi} \right)^{1/2} \left
  ( \frac{\mathcal{A}_i}{\mathcal{A}_i+\mathcal{A}_{li}} 
  \right)^{1/2} =
  \frac{\sqrt{2\sqrt{3}}}{\sqrt{\mathcal{A}_i+\mathcal{A}_{li}}} R_i 
\end{equation}
and 
\begin{equation}\label{PrincenVar2}
\left( \frac{\phi}{\phi_0} \right)^{1/2} \left(\frac{1-\phi_0}{1-\phi}
\right)^{1/2} =
\left(\frac{2\sqrt{3}}{\mathcal{A}_i+\mathcal{A}_{li}}\right)^{1/2}
R_i \left( \frac{2\sqrt{3}
    -\pi}{2\sqrt{3}} \right)^{1/2}
\left(\frac{\mathcal{A}_i+\mathcal{A}_{li}}{\mathcal{A}_{li}} 
\right)^{1/2} = \frac{R_i}{r} 
\end{equation}
where comparison of Eq. \ref{PrincenVar1} with Eq. \ref{PG2} shows
that $(\phi/\phi_0)^{1/2}=P^G_i R_i$. So using Eqs. \ref{PrincenVar1} and
\ref{PrincenVar2}, Eq. \ref{PrincenPi} becomes 
\begin{equation}\label{osmoticP4}
\Pi = \frac{\sigma}{r} - P^G_i
\end{equation}
So since $P_i=P+\sigma/r$, we have the {\sl exact} result that 
$P_i=P+\Pi + P^G_i$, which for the system studied 
confirms both our physical argument for the pressure in an osmotically
compressed bubble, and our hypothesis for this 2D monodisperse system 
that the geometric pressure 
should be calculated for an isotropic expansion at fixed Plateau
border radius $r$ (Eq. \ref{PGhyp}). 

From Eq. \ref{2DGP4}, in the dilute limit of circular bubbles $(r^2 = \mathcal{A}_i/\pi)$
$P^G_i\rightarrow \sigma \sqrt{\pi}/\sqrt{\mathcal{A}_i}$, and in the
dry limit of hexagonal bubbles 
$P^G_i \rightarrow \sigma \sqrt{2\sqrt{3}}/\sqrt{\mathcal{A}_i}$. 
So since $\sqrt{2\sqrt{3}}/\sqrt{\pi} \simeq 1.05$, we find the
surprising result that $P^G_i$ (and hence $P_i$), is only weakly 
affected by $\Pi$. We return to this
in Section \ref{SDF}. 

The area $\mathcal{A}_i^B$ at which bubbles in a hexagonal packing may
coexist with a bulk gas (so that $P_i^s=P+\Pi$), obeys 
$P^G_i(\mathcal{A}_i^B) = P_i^T(\mathcal{A}_i^B)$. So using
Eq. \ref{PG2} and $P_i^T = N^TkT/\mathcal{A}_i$ we obtain  
\begin{equation}\label{2DAB}
\mathcal{A}_i^B = \mathcal{A}^B_{Hi} \left( 1 + \frac{\sqrt{1 +
      {4\mathcal{A}_{li}}/{\mathcal{A}^B_{Hi}}}}{2} 
\right)  
\end{equation}
where $\mathcal{A}_{Hi}^B=(1/4\sqrt{3})(N^TkT/\sigma)^2$, is the area
with which an entirely dry foam with hexagonal bubbles
($\mathcal{A}_{li}=0$) may coexist with a bulk gas. We then obtain the
number of soluble species at
coexistence from $N_i^{sB} = ({\tilde{P}}/{kT}) \mathcal{A}_i^B$.

\section{A Condition for Stability}\label{stabfoamcond} 

Having established a general definition of geometric pressure
(Eq. \ref{PGdef}), and shown that it corresponds (in at least one
special case) to an isotropic expansion at fixed liquid content, 
we now use conservation of the total number of gas molecules and total
number of bubbles to derive a criterion to ensure the formation of a
stable distribution of foam bubbles. 
We take the bubble size distribution to be composed of two
parts, a `coarsening' part of the distribution, and a `stable' part  
consisting of bubbles which have shrunk to a stable size at which they
coexist with the coarsening bubbles.
For a sufficient quantity of trapped species 
the assumption of the bubble distribution having a coarsening part 
is found to be inconsistent, enabling us to derive a stability
criterion for foams.  These arguments have strong similarities to those
in \cite{Webster}, but are considerably generalised. 

We take
$n_b^0$, $n_b^S$, and $n_b^C$ as the number densities of bubbles overall, in the stable part of the distribution, and in the
coarsening part of the distribution respectively.
Conservation of bubble number (no coalescence) gives
\begin{equation}\label{consbub1}
n_b^0  = n_b^S + n_b^C
\end{equation}
where we note that $n^S_b$ and $n^C_b$ may be time-dependent. 
$N_i^{sB}$, the number of soluble species present in bubble $i$ when 
coexisting with a bulk gas phase, is determined by $P_i^G =
P_i^T$. However since relations between $P_i^G$ and $V_i$ may vary
during coarsening (due to changes in the bubble's environment),
we take $N_i^{sB}$ as time-dependent. We define the following
\begin{itemize}
\item{$\bar{N}^{s0} \equiv \langle N_i^{s}(t) \rangle_i$ : The average
    number of soluble molecules per bubble. This is {\sl time
      independent}.} 
\item{$\bar{N}^{sB}(t)$ : The average number of soluble molecules per 
    bubble in the stable distribution. As coarsening proceeds,
    $n^C_b/n^0_b \rightarrow 0$ and the bubbles in the stable distribution
have $N_i^s \rightarrow N_i^{sB}$. Then $\bar N^{sB}(t)$ tends toward the
average number of soluble molecules per bubble at coexistence with a bulk
gas phase. }
\item{$\bar{N}^{sC}(t)$ : The average number of soluble molecules per
    bubble, within the coarsening distribution, at
    time $t$.}  
\end{itemize}

As a coarsening foam tends towards an equilibrium state, 
conservation of the number of soluble molecules requires 
\begin{equation}\label{consmol1} 
n^0_b \bar{N}^{s0} = n^S_b \bar{N}^{sB}(t) + n^C_b \bar{N}^{sC}(t)   
\end{equation}
Combining Eqs. \ref{consbub1} and \ref{consmol1}, gives
\begin{equation}\label{coarseningvsstable} 
n_b^0 \left( \bar{N}^{s0} - \bar{N}^{sB}(t) \right) = n_b^C \left(
\bar{N}^{sC}(t)  - \bar{N}^{sB}(t) \right) 
\end{equation}
So if
\begin{equation}\label{tdepfstab} 
\bar{N}^{sB}(t) \geq \bar{N}^{s0}
\end{equation} 
then Eq. \ref{coarseningvsstable} requires 
$0 \geq n_b^C(\bar{N}^{sC}(t) - \bar{N}^{sB}(t))$. But since the larger,
coarsening bubbles have $\bar{N}^{sC}(t)>\bar{N}^{sB}(t)$, then 
$n_b^C=0$, and the foam must be stable against coarsening. 
Since $\bar{N}^{sB}(t)$ depends on $P^G_i(t)$, whose
relation to $V_i$ is not known, the derivation of a
stability condition in closed form is not always possible, unlike the dilute
case \cite{Webster}. However, the above condition enables us to
investigate the requirements for stability. This is done in
Sections \ref{Exact} and \ref{SDF} below.  

\subsection{Exact Stability Conditions}\label{Exact}

For both dilute foams and the model of a monodisperse 2D foam, the
geometric pressures are calculable exactly, giving   
exact expressions for $N^{sB}_i$. These are:
Eq. \ref{NsB3Ddil}; Eq. \ref{NsB2Ddil}; and
$N_i^{sB}=(\tilde{P}/kT)\mathcal{A}_i^B$ with $\mathcal{A}_i^B$ given
by Eq. \ref{2DAB}; for dilute 3D
foams, dilute 2D foams, and the
monodisperse 2D model respectively. 
Stability conditions are calculated by averaging over the
relevant equation for $N_i^{sB}$ (where appropriate), and ensuring
that Eq. \ref{tdepfstab} is satisfied. 

For dilute 3D foams the requirement is $\bar{N}^{s0} \leq 
\bar{N}^{sB}$, with Eq. \ref{NsB3Ddil} giving 
 \begin{equation}\label{dilcond}
\bar{N}^{sB} = \frac{P}{kT} \sqrt{ \frac{3}{4\pi}} \left
  ( \frac{kT}{2\sigma} \right)^{3/2} \langle {N_i^T}^{3/2} \rangle_i 
\end{equation}
which resembles, but generalises a result in \cite{Webster}.
Similarly for dilute 2D foams we require $\bar{N}^{s0} \leq 
\bar{N}^{sB}$, but with Eq. \ref{NsB2Ddil} giving     
\begin{equation}\label{2Ddilcond}
\bar{N}^{sB} = \frac{P}{kT} \left( \frac{1}{\pi} \right)
\left( \frac{kT}{\sigma} \right)^2 \langle {N_i^T}^2 \rangle_i 
\end{equation}
Finally, since the model of 2D foams considers monodisperse bubbles we 
merely require that $N^{s0}_i \leq N^{sB}_i$, with Eq. \ref{2DAB} 
giving 
\begin{equation}\label{NsB2D}
N^{sB}_i = \frac{\tilde{P}}{kT} \mathcal{A}_i^B = \frac{\tilde{P}}{kT}
\mathcal{A}^B_{Hi} \left( 1 + \frac{\sqrt{1 +
  {4\mathcal{A}_{li}}/{\mathcal{A}^B_{Hi}}}}{2} \right)  
\end{equation}
Recall that $\mathcal{A}_{Hi}^B=(1/4\sqrt{3})(N^TkT/\sigma)^2$ so that $N_i^{sB}$ depends on $\sigma$ as expected. The area of liquid associated
with each bubble becomes negligible compared to $\mathcal{A}_{Hi}^B$ as the foam becomes increasingly dry. Note also that the dry monodisperse hexagonal foam is,
even without trapped species, already 
dynamically stable with respect to infinitesimal volume changes of a single cell but not with
respect to geometric reorganisation so as to create five-sided and seven-sided
cells. Such a foam is, however, unstable with respect to homogenous cell shrinkage
throughout the system (with the excess gas forming a bulk coexisiting phase), whereas sufficient trapped species, as calculated above, will restore full thermodynamic stability in this sense. We accept that the distinction
between stability and metastability becomes blurred, especially as the geometry of the foam becomes more complex.

\section{Polydisperse Foams}\label{2Dpolysystems} 

The edges of bubbles in a reasonably dry 2D foam ($\Pi\gg
\sigma/R_i$), meet with an angle approximately equal to $2\pi/3$. 
If the edges met with an angle of {\sl exactly} $2\pi/3$, then bubbles
would need to have {\sl exactly equal} radii of curvature at their
Plateau borders, and hence {\sl equal} bubble pressures
$P_i=P+\sigma/r$. Since the curvatures between adjacent bubbles remain 
of order $1/R_i \ll 1/r_i$, then as a foam becomes increasingly dry,  
bubbles' pressure differences become increasingly small compared with
the mean bubble pressure ($P_i \gg P_i^G \sim \sigma/R_i$).
So provided a polydisperse foam is sufficiently
dry, then $P_i^G\sim \sigma/R_i$ and $P_i^G\ll \Pi$.    

\begin{figure}[htd]
\begin{center}
\leavevmode
\epsfig{file=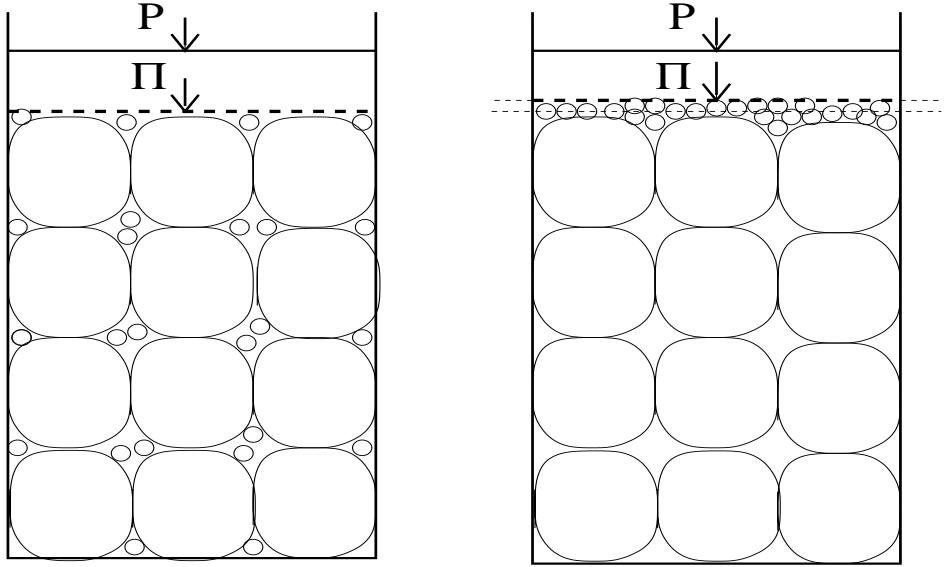,width=12.5cm,height=7.5cm}    
\end{center}
\caption{At a given osmotic pressure $\Pi$, the volume occupied by
  a weakly compressed polydisperse foam can depend on the way in
  which bubbles are arranged. The two foams shown 
  both contain the same
  bubbles but the different bubble arrangements result in their
  occupying different volumes.}\label{voldependence}  
\end{figure}

In a sufficiently {\sl wet} and polydisperse foam, 
very small bubbles might reside within the 
Plateau borders of larger bubbles without mechanically experiencing an
osmotic pressure $\Pi$. Such a bubble in a 3D foam will be spherical
with $P_i=P+2\sigma/R_i$ and hence $P_i^G=2\sigma/R_i-\Pi$. For such
bubbles $P_i^G$ need not necessarily be of order $\sigma/R_i$.    
In fact even for the same bubble size 
distribution and the same osmotic pressure $\Pi$, a foam's {\sl volume} 
can depend on the arrangement of bubbles it contains (see
figure \ref{voldependence}). 
However, given 
a reasonable osmotic compression $\Pi$, then a reasonably
narrow distribution of $N_i^T$ will be sufficient to ensure that such
bubbles occupy a negligible volume fraction, and 
hence may be neglected when calculating the stability condition
Eq. \ref{tdepfstab}. This is shown in Appendix \ref{polydappendix}.

\subsection{Stability Requirements}\label{SDF} 

\begin{figure}[htd]
\begin{center}
\leavevmode
\epsfig{file=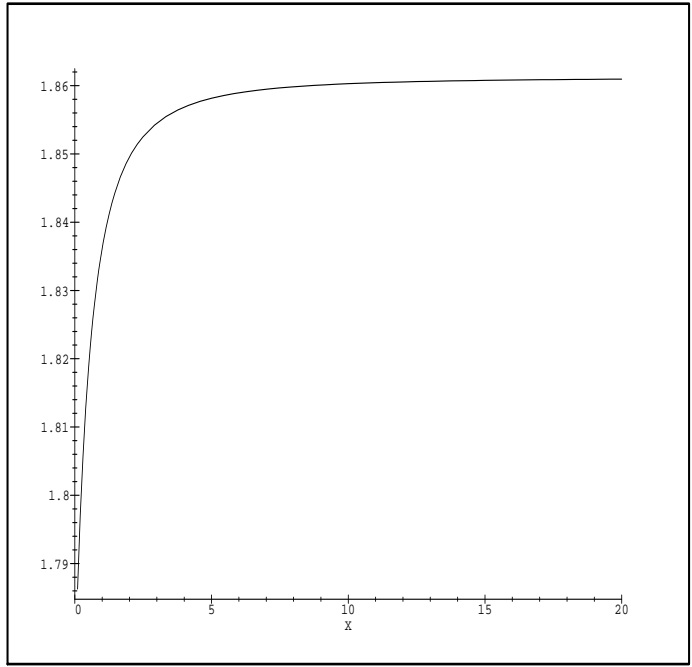,
width=10cm,height=10cm,angle=270}   
\end{center}
\caption{Variation of $\Gamma_i$ (vertical axis), from $\sqrt{\pi}$ to
  $\sqrt{2\sqrt{3}}$,  with $X\equiv
  l/r$. Increasing $X$ corresponds to the bubble becoming increasingly
  dry, and $X \rightarrow 0$ corresponds to the bubble becoming
  circular in shape.}\label{Gammapict}  
\end{figure}

To address the stability condition Eq. \ref{tdepfstab}, we need to
know the size at which shrunken bubbles in the stable
distribution will coexist with a bulk phase of soluble gas.  
We will assume that for a given $\Pi$ the distribution of $N_i^T$ is 
sufficiently narrow that we may neglect any tiny bubbles 
residing wholly within Plateau borders of larger ones (see above, and Appendix \ref{polydappendix}). 
Then for a shrunken bubble in $\mathcal{D}$
dimensions of volume $V_i$, we define a quantity $\Gamma_i$ such that 
\begin{equation}\label{PG} 
P^G_i = \Gamma_i \frac{\sigma}{V_i^{1/\mathcal{D}}} 
\end{equation}
and expect $\Gamma_i \sim 1$. (Any coarsening or rearrangements will make
$\Gamma_i$ time dependent.) 
For example, our model 2D foam has (from Eq. \ref{PG2}) 
\begin{equation}\label{gammacurve}
\Gamma_i = \sqrt{2\sqrt{3}} \sqrt{1 -
  \frac{\mathcal{A}_{li}}{\mathcal{A}_i+\mathcal{A}_{li}} }   
\end{equation}
which is graphed in figure \ref{Gammapict}. 
In this example $\Gamma_i$ varies monotonically 
between $\sqrt{\pi}$ for an entirely
dilute foam (circular bubbles), and $\sqrt{2\sqrt{3}}$ for an entirely
dry foam (hexagonal bubbles), but always remains of order $1$.  

We may calculate an approximate value for the volumes of shrunken 
bubbles by taking $\Gamma_i$ as independent of $V_i$ (although
changes in a bubble's environment mean that $\Gamma_i$ may
change with time). Using $N_i^{sB}=(\tilde{P}/kT)V_i^B$, and
obtaining $V_i^B$ by equating $P_i^G=\Gamma_i\sigma V_i^{1/\mathcal{D}}$ with
$P_i^T=N_i^TkT/V_i$, gives
\begin{equation}\label{ansatzEq}
N_i^{sB} = \frac{\tilde{P}}{kT} \left( \frac{N_i^T kT}{\sigma \Gamma_i}
  \right)^{\mathcal{D}/(\mathcal{D}-1)}  
\end{equation}  

In Appendix \ref{ReqApp} we use Eq. \ref{ansatzEq} to investigate the
stability requirement Eq. \ref{tdepfstab} in more detail. By assuming that
$\Gamma_i$ and $N_i^T$ are uncorrelated, it is shown that unless the
average value $\bar{\Gamma}$ may increase {\sl without bound}, then
a stability threshold does exist. 
(That is, there will always be some number of trapped molecules per
bubble beyond which coarsening will be prevented.) 
Equation \ref{gammacurve} shows that, for a monodisperse 2D foam, 
$\Gamma$ is bounded above even in the dry limit (in contrast to the Laplace pressure). 
Also since $\langle P^G_i \rangle \sim
\sigma/\bar{R}$ (see appendix \ref{Fbalance}), then for $\bar{\Gamma}$ to increase without bound,
$\Gamma_i$ and $R_i$ will need to be correlated in a very specific
way. On balance, all these arguments suggests that
$\bar{\Gamma}$ remains bounded as a polydisperse foam evolves. 
Hence our results suggest it should always be possible
to osmotically stabilise a polydisperse foam by adding enough trapped species.
As a rough estimate, stability requires a sufficient pressure of
trapped gas that $\bar{P}^T\grtapprox \sigma/\bar{R}$, 
which for $\sigma\sim 10^{-1}$Nm$^{-1}$, and $\bar{R}\sim 10^{-5}$m
requires $\bar{P}^T\sim 10^{-4}$Nm$^{-2}$, ie of order $0.1$ of 
atmospheric pressure.


\section{Coarsening of Foams: Qualitative Behaviour}\label{CoarsenFchapt} 

Now we consider the coarsening of incompletely stabilised, non-dilute
foams. Previous work on coarsening of dilute emulsions 
in \cite{Webster} also applies to dilute foams {\sl mutatis mutandis}. 
Here we concentrate on non-dilute foams
in which bubbles impinge on one another and are distorted from their
otherwise spherical shape. 

As with dilute emulsions\cite{Webster}, the trapped molecules  
prevent bubbles from entirely disappearing. The resulting foam
morphology and coarsening kinetics will be determined by 
two main factors. The first is the `excess volume fraction' of dispersed
phase, defined as the total volume fraction of gas which ultimately
will coexist with a stable ensemble of shrunken bubbles, in osmotic
and mechanical equilibrium with it: this is the amount of gas 
actually available for coarsening. The second factor is the ease with which
shrunken bubbles may rearrange to allow larger bubbles to coarsen.  

\begin{figure}

\noindent

\begin{minipage}[b]{.3\linewidth}
\centering
\epsfig{file=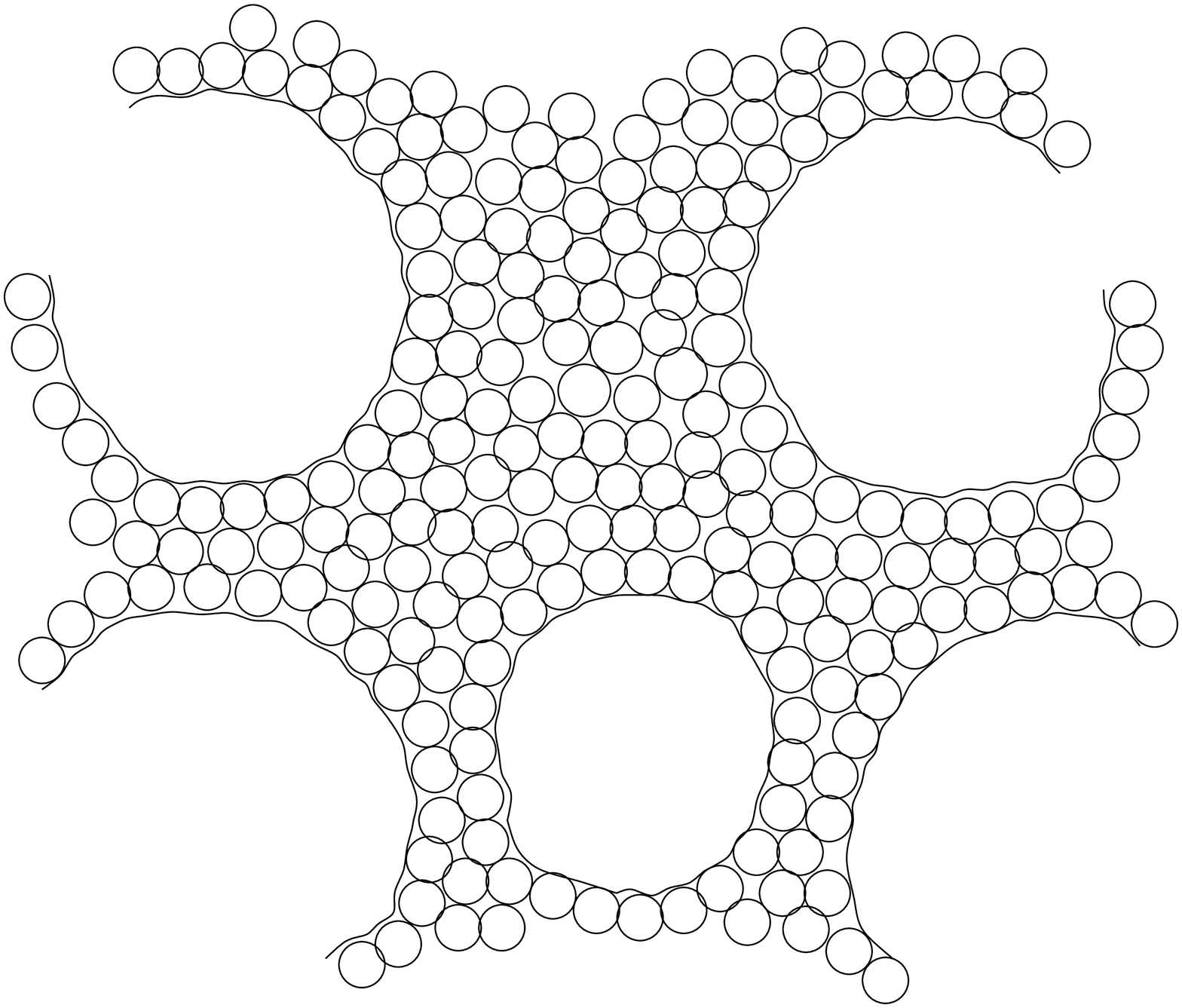,width=4.5cm, height=5cm}
\label{lowexcess} 
\end{minipage}
\hfill 
\begin{minipage}[b]{.3\linewidth}
\centering
\epsfig{file=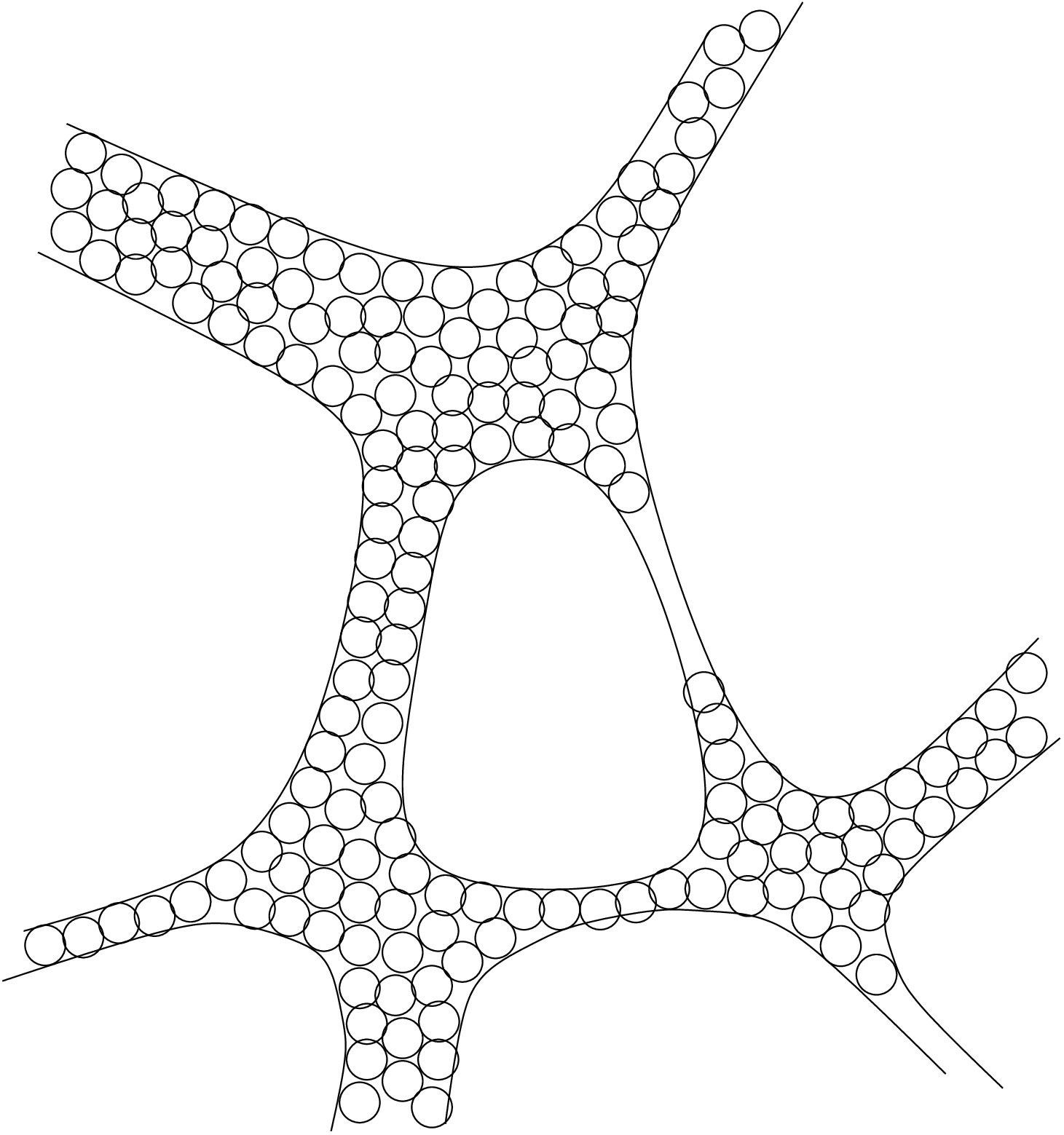,width=4.5cm,height=5cm,angle=0}    
\end{minipage} 
\hfill
\begin{minipage}[b]{.3\linewidth}
\centering
\epsfig{file=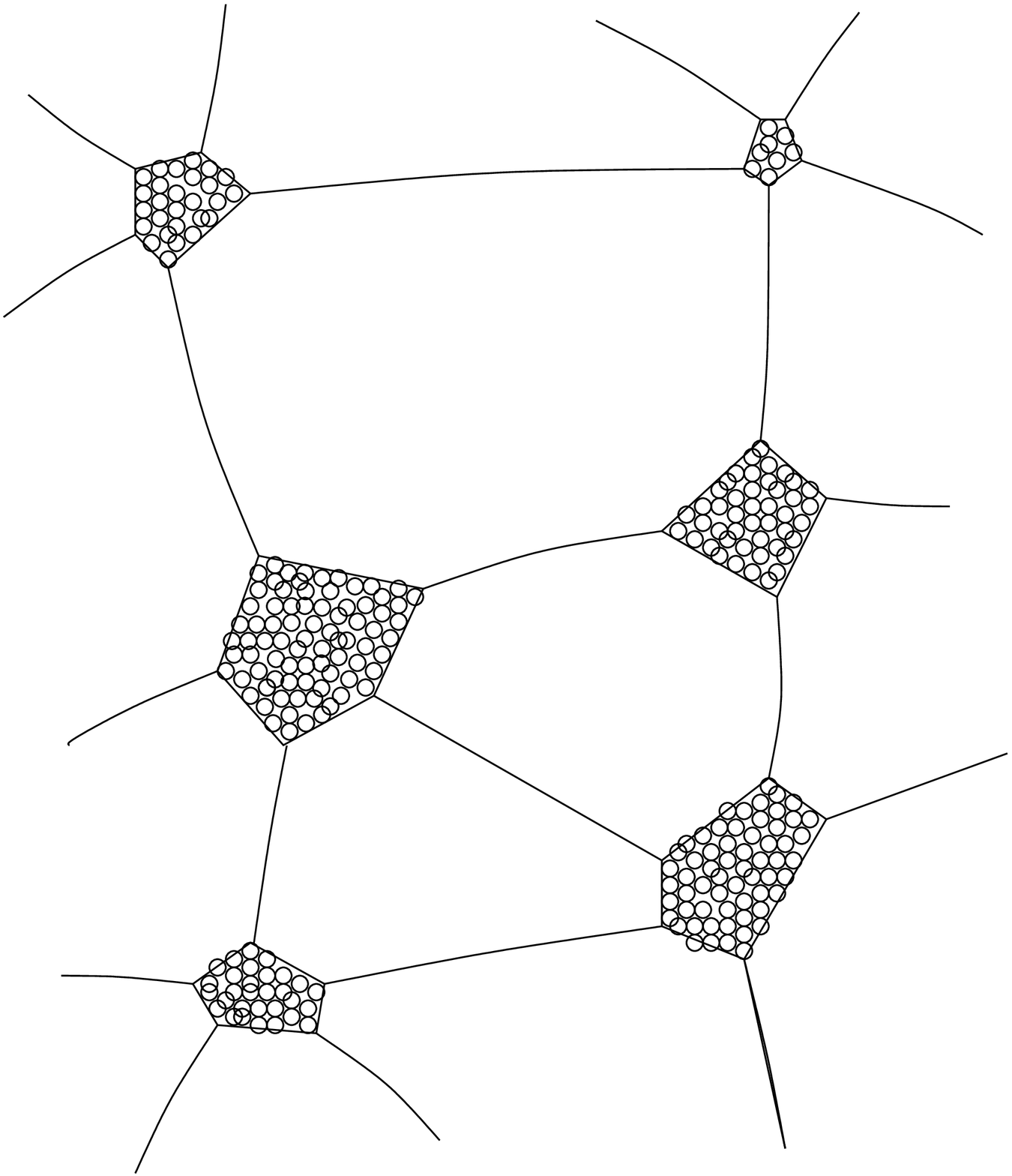,width=4.5cm,
  height=5cm}   
\label{highexcess} 
\end{minipage}

\caption{Late stage foam morphologies. For low excess volume fractions (left)
  grown bubbles are well separated by large numbers of shrunken bubbles,
  whereas at high excess volume fractions (right) the grown bubbles impinge
  on one another with shrunken bubbles decorating the grown bubbles'
  vertices.}\label{excessfig}
\end{figure}

The excess volume fraction
of disperse phase, determines a foam's expected
late-stage morphology (see figure \ref{excessfig}): 
\begin{enumerate}
\item{{\sl No excess volume fraction:} The foam is stable.}
\item{{\sl Very low excess volume fractions:} 
    Larger bubbles are surrounded by a `sea' of shrunken bubbles. 
    Competitive coarsening between larger bubbles requires a gas
    flux {\sl through} the sea of smaller bubbles.}
\item{{\sl Very high excess volume fraction:} 
    Larger bubbles are {\sl decorated} by collections of
    smaller bubbles at their {\sl vertices}, with their
    {\sl faces} impinging on other large bubbles.} 
\item{{\sl Intermediate volume fractions:}
      Structures between the previous two extremes.} 
\end{enumerate}

In what follows, we focus on the case of low but nonzero excess volume fraction. Here coarsening of larger bubbles will require rearrangements of the
shrunken bubbles to prevent the build up of excess elastic strains (which
will otherwise halt coarsening, as shown below). 
We envisage four scenarios: 
\begin{enumerate}
\item{{\sl Inviscid rearrangements:} Bubble rearrangements occur
    easily, and with negligible dissipation of energy.}
\item{{\sl Viscous rearrangements:} Bubbles can rearrange,
    but resist doing so and hence slow the rate of
    coarsening.}
\item{{\sl Negligible rearrangements (elastic medium):} Bubbles grow within an
    effectively elastic medium, which may eventually arrest coarsening.}
\item{{\sl Elasto--plastic rearrangements:} There is a maximum yield
strain beyond which rearrangements allow flow of the shrunken--bubble sea, but below which it behaves as an elastic medium.}
\end{enumerate}
We expect rearrangements to occur easily in a sufficiently wet foam,  
but rearrangements in a very dry foam to occur rarely or not
at all. So we expect the scenarios from $1 \rightarrow 3$, to become more
applicable as foams become increasingly dry. 
Since {\sl both} rearrangements and diffusion of disperse
phase are required for coarsening to occur, coarsening
will proceed with a rate determined by the {\sl slowest} process. 
This contrasts, for example, with phase separation in a binary fluid, where the 
coarsening rate is governed by the {\sl fastest} process (with
diffusive coarsening at early times, viscous hydrodynamic
coarsening at intermediate times, and inertial hydrodynamic
coarsening at late times \cite{Bray}). 
In the present case rapid coarsening may initially be limited by
viscous forces, then later as coarsening slows, viscous forces will  
become negligible and coarsening diffusion-limited. 

Note that a similar classification of kinetic regimes may in part be applicable
to the coarsening dynamics of foams containing no trapped species. It would be appealing, perhaps, to study this case in detail
first, before addressing the situation where trapped species
are present. However, the latter case is actually a lot simpler,
at least in the case of low excess volume fractions (figure \ref{excessfig}, left) considered here. This is because the actively coarsening bubbles are
effectively decoupled by a sea of passive shrunken bubbles; this simplification
is absent at excess volume fractions approaching unity (which recovers the unstabilised case).

\section{Mean-Field Model: Inviscid Rearrangement}\label{rapidR}

We consider a mean-field model for the coarsening of a small excess
volume fraction, in which grown bubbles
are already sufficiently large that they contain a negligible quantity
of trapped molecules, and have an approximately spherical shape 
(figure \ref{lowexcess}).
Hence large bubbles have $P_i^G \simeq 2\sigma/R_i$ and $P_i^S \simeq 
\tilde{P} + 2\sigma/R_i$.   
We restrict ourselves to incompressible foams (in the sense described
in Section \ref{compressed/drained}), and take shrunken bubbles to
have an approximately constant size $V_B \equiv {4\pi}
{R_B}^{3}/3$, with $R_B \ll R_i$.  
An average pressure $P^s(r,t)$ of soluble gas in shrunken bubbles at distance $r$ 
from a grown bubble's centre is obtained by coarse-graining over
bubbles at $r$, and we assume $P^s(\infty,t)$ is the same for all
grown bubbles.  
We at first take rearrangements to be inviscid, so that bubble growth
is determined by the rate at which soluble gas diffuses through
bubble-bubble interfaces in the shrunken bubble `sea'; this assumption is
relaxed in Section \ref{DissipRates}.

\begin{figure}[htd]
\begin{center}
\leavevmode
\epsfig{file=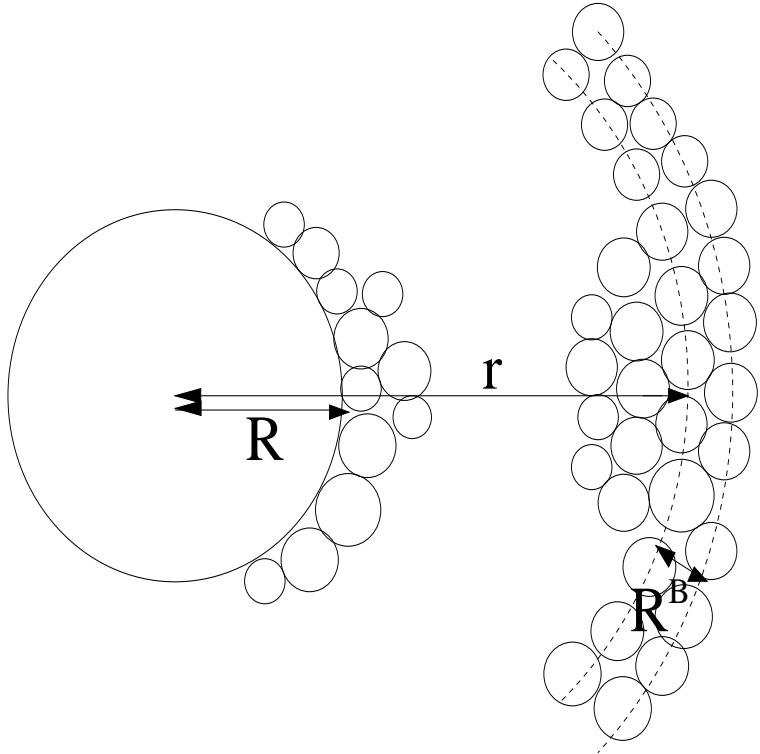,
width=10cm,height=10cm,angle=270}   
\end{center}
\caption{We consider a grown bubble with $R \gg R^B$, and shrunken
bubbles at a distance $r$ from its centre to have an
average pressure of soluble gas $P^s(r,t)$.}\label{continuum_sea}
\end{figure} 

Consider the flux of gas from shrunken bubbles at radius $r$ to
adjacent bubbles at radius $(r+ R^B)$, see figure \ref{continuum_sea}. 
Following the approach of Von Neumann (see \cite{Glazier}), we take
the flux of gas between two bubbles as proportional to both the pressure
difference of their soluble gas, and the surface area
through which the gas may pass. 
Defining a flux velocity per unit pressure $K$, we obtain an average 
volume flux of gas from bubbles at $r$ to bubbles at $(r+R^B)$ of   
\begin{equation}
\mathbf{J}_V(r,t) = K 4 \pi r^2 \left(P^s(r,t) - P^s(r+R_B,t)
  \right) \mathbf{\hat{r}}
\end{equation} 
Since $R\gg R^B$ then $(P^s(r,t)-P^s(r+R_B,t))/R_B\simeq \partial
P^s(r,t)/\partial r$, so solving for a steady state gas flux with
$\mathbf{\nabla} . \mathbf{J} = 0$, we obtain a radial flux
\begin{equation}
{J}_V(R,t) = K 4 \pi R^2 R^B \frac{P^s(R)-P^s(\infty,t)}{R}
\end{equation} 
and a droplet growth rate  
\begin{equation}
\frac{dR}{dt} = \frac{KR^B}{R} \left( \Delta P^s(\infty,t) -
  \frac{2\sigma}{R} \right) 
\end{equation}
where we used $P^s(R) = \tilde{P} + {2\sigma}/{R}$, and wrote 
$\Delta P^s(\infty,t) = P^s(\infty,t) - \tilde{P}$. 
For an incompressible system all bubbles have a volume per gas
molecule equal to that of a bulk gas, $v_g \equiv {kT}/{\tilde{P}}$.
So with a little algebra we get 
\begin{equation}\label{eff_growthrate} 
\frac{dR}{dt} = \frac{K\tilde{P}R_B}{R} \left( 
\epsilon - \frac{2\sigma v_g}{kTR} \right) 
\end{equation}
where $\epsilon \equiv 
(P^s(\infty,t) - \tilde{P})/\tilde{P}$. 

Previously the thickness of liquid films between bubble faces was
taken as zero. Now we take such films to have a small but
finite thickness $d$. Then assuming the rate of flux through liquid
films to be diffusion limited, we may calculate $K$ in terms of the
diffusion constant for dissolved gas molecules $D$, the volume per gas
molecule $v_g$, $\tilde{P}=P+\Pi$, and $d$. 
This gives\cite{WebsterTh}  
\begin{equation}
K = \frac{Dv_gC(\infty)}{\tilde{P}d} 
\end{equation}
So the growth rate may then be written as
\begin{equation}\label{RRgrowrate}
\frac{dR}{dt} = \left( \frac{R_B}{d} \right) 
\left( \frac{Dv_gC(\infty)}{R} \right)
\left( \epsilon  - \frac{2\sigma v_g}{kTR} \right)  
\end{equation}
Note that for a {\sl dilute} foam\cite{WebsterTh} the factor $R_B/d$
is absent from Eq. \ref{RRgrowrate}. 
The increase in growth rate is due to the reduced volume fraction of
liquid through which gas molecules actually need diffuse, a reduction of order 
$d/R_B$.

\section{Rate-Limiting Mechanisms}\label{DissipRates}

We continue to study the mean-field model of Section \ref{rapidR},
but no longer require inviscid rearrangements. 
When bubble rearrangements {\sl are} inviscid there is dissipation
from the {\sl diffusion resistance} of dissolved-molecules diffusing
through liquid films between bubbles,
and if bubble rearrangements are viscous there is also dissipation as
bubbles rearrange.
By equating the rate of dissipation with the rate of decrease in free
energy, in Sections \ref{LSDissip}, \ref{inviscidDissip}, and
\ref{viscousDissip} we obtain the order of magnitude for a droplet's
growth rate. 
The {\sl greatest} source of dissipation limits the coarsening rate,
so by comparing the dissipation rates we can estimate (Section \ref{viscousDissip}) when each type of coarsening will occur. 
A similar approach applied to emulsion rheology (not coarsening), is
found in \cite{BuzzaCates}.  

As an example, we firstly consider coarsening 
in the traditional LSW\cite{Lifshitz,Wagner} scenario
of a vanishingly  small volume fraction of bubbles in a liquid
(without trapped molecules). Here dissipation is from the diffusion
resistance of diffusing, dissolved gas molecules. 

\subsection{LSW Coarsening of Bubbles}\label{LSDissip}

In a steady state the radial flux of dissolved gas molecules, at a distance $r$ from the centre of a bubble of
radius $R$ is 
\begin{equation}
{J}(r) = \frac{4\pi}{v_g} \frac{R^2\dot{R}}{r^2}
\end{equation}  
We write ${J}(r)=c(r)u(r)$, with $c(r)$ and
$u(r)$, the concentration 
and average radial velocity of dissolved gas molecules at $r$
respectively. Since $c(r) \sim C(\infty)$, where $C(\infty)$ is
the concentration of dissolved gas molecules adjacent to a bulk gas
phase, then 
\begin{equation}\label{diffvel}
u(r) \sim \frac{R^2\dot{R}}{v_gC(\infty)} \frac{1}{r^2} 
\end{equation}

The number of dissolved gas molecules within spheres of radii $r$
and $r+dr$ is of order $r^2 dr c(r) \sim r^2 dr C(\infty)$.
The average dissipation rate per molecule at $r$ is $\zeta
u(r)^2$, where $\zeta$ is the viscous drag coefficient on a molecule
of disperse phase moving through the liquid.  
Hence the total dissipation arising from diffusion to a drop is of
order $\int^{\infty}_R r^2dr C(\infty) \zeta u(r)^2$.
Using the Einstein relation\cite{Reif} $\zeta = kT/D$,
Eq. \ref{diffvel}, and integrating, we obtain the rate of dissipation
due to dissolved molecules diffusing between bubbles 
$T\dot{S}_D$, as 
\begin{equation}\label{dissipation1}
T\dot{S}_D \sim \frac{kT}{DC(\infty)v_g^2} \dot{R}^2 R^3 
\end{equation}

Coarsening occurs to reduce interfacial energy, with a rate of
reduction in energy of order $\sigma R\dot{R}$. So equating $\sigma
R\dot{R}$ with Eq. \ref{dissipation1} and rearranging, we get 
\begin{equation}
\frac{dR}{dt} \sim \frac{Dv_g^2 C(\infty)\sigma}{kTR^2} 
\end{equation}
in agreement with the traditional analysis of LSW\cite{Lifshitz,Wagner} 
for coarsening of dilute emulsion droplets (as opposed to foam bubbles
as studied here).

\subsection{Inviscid Rearrangements Revisited}\label{inviscidDissip} 

We now apply the method to the mean-field model of Section
\ref{rapidR}, where inviscid rearrangements were assumed. 
The argument in Section \ref{LSDissip} gives the average velocity with 
which dissolved gas molecules will diffuse through liquid 
films, in Eq. \ref{diffvel}, and the dissipation per molecule in
liquid films remains $\zeta u(r)^2$.  
But now dissipation only occurs in the liquid films between bubble
faces. To integrate only over the volume of such films,  
the spatial volume element $r^2dr$ is reduced by a factor of $d/R_B$. 
So the dissipation due to diffusion resistance $T\dot{S}_D$ is 
\begin{equation}\label{invisciddissip}
T\dot{S}_{D} \sim \frac{d}{R_B} \frac{kT}{DC(\infty) v_g^2} \dot{R}^2 
R^3  
\end{equation}
and we obtain 
\begin{equation}\label{mce10.22}
\frac{dR}{dt} \sim \left(\frac{R_B}{d}\right) \left( \frac{Dv_g
    C(\infty)}{R} \right) \left( \frac{ \sigma v_g}{kTR} \right)
\end{equation}
in agreement with the mean-field calculation (Eq. \ref{RRgrowrate}).
This gives a growth law for the mean droplet size, $\bar{R}\sim t^{1/3}$.

\subsection{Viscous Rearrangements}\label{viscousDissip}

We now consider the rate of dissipation due to viscous stresses in
the thin liquid films, as shrunken bubbles rearrange. 
Adjacent bubble faces are again taken to be separated by a distance $d$,
determined by the disjoining pressure between bubble membranes.
Since the excess volume fraction is small, we assume that bubble
growth results in fluid flow that is approximately radial.
Taking $\mathbf{v}(\mathbf{r})$ as the velocity of the 
shrunken-bubble fluid, then incompressibility of that fluid 
requires ${\nabla} . \mathbf{v}(\mathbf{r}) =0$, so that in
the spherically symmetric case there is a radial velocity 
\begin{equation}\label{rfluidvel}
v(r) = \frac{R^2\dot{R}}{r^2} 
\end{equation}
at distance $r$ from the centre of a bubble of radius $R$, 
growing with velocity $\dot{R}$.

Consider now two adjacent shrunken bubbles at distances $r$ and $r+R_B$
respectively from the centre of the growing bubble. 
The differing velocities at $r$ and $r+R_B$ will mean
that bubbles must rearrange, and slide past one another. For $r \gg
R_B$ the relative velocity of the bubbles is 
of order $ ({\partial v(r)}/{\partial r}) R_B$. 
The shear rate of the liquid between bubbles is of the order of this
relative velocity divided by the
film thickness $d$. Within the film the viscous stress is therefore  
\begin{equation}
\eta\dot \gamma (r)\sim \eta \frac{R_B}{d} \frac{\partial v(r)}{\partial r}
\end{equation}
where $\eta$ is the viscosity of the continuous liquid phase, and
$\dot{\gamma}(r)$ is the shear rate in a liquid film at $r$.   

The volume-averaged viscous dissipation is 
dominated by the contribution from within the films \cite{BuzzaCates}
and, per unit volume, is of order $d/R_B$ times
the dissipation rate (of order $\eta\dot{\gamma}(r)^2$) within each film. 
So we obtain the total dissipation rate due to viscous bubble
rearrangements, $T\dot{S}_V$, as 
\begin{equation}
T \dot{S}_{V} \sim \int_R^{\infty} r^2dr \frac{d}{R_B} \eta \left(  
\frac{R_B}{d}   \frac{\partial v(r)}{\partial r} \right)^2 
\end{equation}
Using Eq. \ref{rfluidvel} and integrating, this gives 
\begin{equation}\label{viscousdissip}
T \dot{S}_{V} \sim \frac{R_B}{d}\eta  R\dot{R}^2  
\end{equation}

As in the previous calculations (Sections \ref{LSDissip} and
\ref{inviscidDissip}), the rate of decrease in the surface free energy of
the growing bubble is of order  
$\sigma R \dot{R}$. Equating $\sigma R \dot{R}$ with
Eq. \ref{viscousdissip}, and rearranging gives
\begin{equation}\label{mce10.30}
\dot{R} \sim \frac{\sigma}{\eta} \frac{d}{R_B} 
\end{equation}
and hence a linear growth law, $\bar{R}\sim t^1$. 

It is interesting to ask why the above argument and the resulting linear growth law does {\sl not} apply to the conventional LSW coarsening of dilute emulsion
droplets in a structureless fluid continuum. In emulsions the volume of a molecule in a droplet is similar
to that in the continuous phase, so when a dissolved molecule is
incorporated into a drop, the increase in the  droplet's volume equals
the volume of liquid previously displaced by the molecule. 
Hence the only displacement of liquid is that already accounted for
by the Stokes-Einstein drag on the molecule as it diffuses through the
continuous liquid phase. On the other hand, since the 
volume per gas molecule in
a bubble is much larger than its volume when dissolved in solution,
the viscous dissipation {\sl may} be relevant for coarsening of
dilute foam bubbles in a structureless fluid. (In this case, arrival of gas 
molecules at the surface of a growing bubble causes a net fluid flow 
radially outward.) A simple order of
magnitude estimate gives 
\begin{equation}\label{dilTSdotV}
T\dot{S}_V \sim \eta R\dot{R}^2
\end{equation}
and $\dot{R}\sim \sigma/\eta$. This differs from the case where the surrounding medium is a `shrunken bubble' fluid by the absence of the enhancement
factor $R_B/d$.

\subsection{Viscous or Diffusion Limited Growth?}\label{VorDltd}

In a coarsening dense foam, both rearrangements and diffusion are necessary for bubble
growth, so the growth rate will be limited by the {\sl greatest} source of
dissipation.
So when $T\dot{S}_{D}/T\dot{S}_{V}\gg 1$ coarsening
is diffusion limited, and when $T\dot{S}_{D}/T\dot{S}_{V}\ll 1$
coarsening is limited by viscous dissipation.  Here we estimate the ratio for
plausible foam parameters.

Comparing Eqs. \ref{invisciddissip} and \ref{viscousdissip}, we see
that  
\begin{equation}
\frac{T\dot{S}_{D}}{T\dot{S}_{V}} \sim \left( \frac{d}{R_B}
\right)^2 \frac{kT}{DC(\infty) v_g^2 \eta} R^2 
\end{equation}
In terms of a molar concentration $C_M$, 
the molar volume $v_{Mg}$ of gas in bubbles at pressure $\tilde{P}$,
and the gas constant $R_G$, we have  
\begin{equation}\label{aw10.34}
\frac{T\dot{S}_{D}}{T\dot{S}_{V}} \sim \left( \frac{d}{R_B}
\right)^2 \frac{R_GT}{DC_M(\infty) v_{Mg}^2 \eta} R^2 
\end{equation}
At room temperature and atmospheric pressure we take typical values of 
$R_GT \sim 10^3$J, $D\sim 10^{-9}$m$^2s^{-1}$,
$C_M(\infty)\sim 10^2$m$^{-3}$, $v_{Mg} \sim 10^{-2}$m$^3$, and $\eta
\sim 10^{-3}$Nm$^{-2}$s,  
(eg. for CO$_2$ gas bubbles in water \cite{CRC}). 
Taking $R\sim 10^{-6}$m, gives  
\begin{equation}
\frac{T\dot{S}_{D}}{T\dot{S}_{V}} \sim 10^{5}  \left
  ( \frac{d}{R_B} \right)^2 
\end{equation}
So that for micron-sized foam bubbles, viscous dissipation will be
observed for $d/R_B$ smaller than $10^{-2.5}$. Thus for sufficiently
small bubbles, thin liquid films, and high liquid viscosity, 
viscous limited growth may be observed (giving $\bar{R}\sim t$).  
However at room temperature and pressure we expect diffusion
limited coarsening to be more common, and moreover this will always
dominate once the radius $R$ of the coarsening droplets becomes sufficiently
large.
Note that for foams $v_{Mg}=R_GT/\tilde{P}$, so
$T\dot{S}_D/T\dot{S}_V$ is proportional to $\tilde{P}^2/kT$:
at low pressures and high temperatures the prospect of
observing viscous limited coarsening is increased.   
For example if $\tilde{P}\sim 10^3$Nm$^{-2}$ ($10^{-2}$ atmosphere)
then $T\dot{S}_D/T\dot{S}_V \sim
10 (d/R_B)^2$, which for $d/R_B\sim 10^{-1}$ predicts viscous limited
growth. 

For dilute foams $T\dot{S}_D$ and $T\dot{S}_V$ are given by
Eqs. \ref{dissipation1} and \ref{dilTSdotV} respectively, so that
\begin{equation}
\frac{T\dot{S}_{D}}{T\dot{S}_{V}} \sim 
\frac{R_GT}{DC_M(\infty) v_{Mg}^2 \eta} R^2 
\end{equation}
Hence at room temperature and pressure, taking $R_GT$, $D$,
$C_M(\infty)$, $\eta$, $v_{Mg}$, and $R$ as above,
$T\dot{S}_D/T\dot{S}_V \sim 10^5$ and coarsening will be diffusion
limited\cite{ff1}.
Writing $v_{Mg}=R_GT/\tilde{P}$ we find that 
viscous limited growth requires $\tilde{P}^2/R_GT\lessapprox 10^2$Nm
which at room temperature requires $\tilde{P} \lessapprox 10^{-3}$ atmosphere.

\section{Negligible Rearrangements: Elastic Medium}\label{elasticD}

Suppose we no longer allow bubbles to rearrange, so that the sea of
shrunken bubbles acts as an elastic medium. By estimating the
increase in elastic energy as a growing bubble changes volume, we show
that coarsening will halt, with the foam now `elastically' (as opposed
to `osmotically') stabilised. 

The assumption of negligible rearrangement requires detailed explanation,
since in any foam (containing trapped species or otherwise) local rearrangement follows inevitable when certain conditions are reached at the junctions
between adjacent cells. However, it is well known that, {\em in the absence of
coarsening}, a foam can exhibit a finite macroscopic yield strain, below which
there may be occasional local rearrangements but these are insufficient to
allow plastic deformation of the medium as a whole \cite{Weaire_Pageron}.

Our physical picture is that the shrunken--bubble sea, which is fully stabilised
against coarsening by virtue of the trapped species that it contains, can then offer elastic resistance to the growth of an isolated large bubble, even though
the latter is at lower chemical potential of gas than other, less large bubbles that have coarsened elsewhere in the system.

The elastic coarsening inhibition mechanism that we advance is thus only possible because of the stability against coarsening of the shrunken--bubble sea. If the bubbles in this sea were themselves members of the coarsening
population, then the rearrangement conditions at the junctions between them
would be constantly met within it, and there would be an incessant
rearrangement of bubbles caused by their local volume changes. Such a medium could not exhibit a yield stress in any conventional sense, and would be unable to elastically stabilise the growth of isolated large bubbles. This remark is consistent with von Neumann's theorem, which (in two dimensions) establishes
that in the absence of trapped species, coarsening can never cease for any structure other than a perfect hexagonal lattice. The theorem does not apply in 
the presence of trapped species \cite{vonfootnote}, and so the elastic arrest
mechanism considered below leads to no contradiction with it.

For definiteness we consider an initial state comprised of
spherical {\sl ready-grown} bubbles in an elastically unstrained sea of shrunken
bubbles (see figure \ref{excessfig}).  
Such a system could be formed by osmotically compressing a previously
dilute, partially coarsened foam. 

\begin{figure}[htb]
\begin{center}
\leavevmode
\epsfig{file=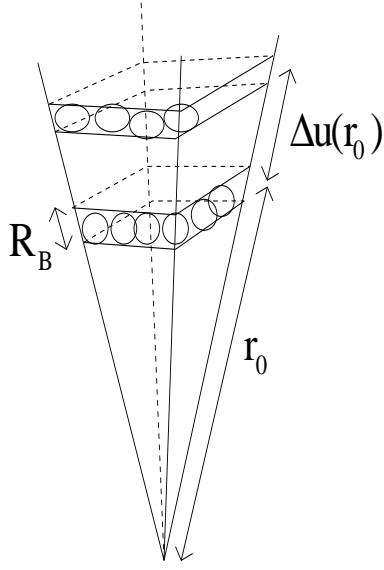,width=5cm,height=7.5cm}    
\end{center}
\caption{In the absence of rearrangements, bubble growth
  results in an increase in the shrunken bubbles' interfacial
  area.}\label{bubblelayer} 
\end{figure}  

Consider a layer of shrunken
bubbles initially at distance $r_0$ from the centre of a grown bubble
of initial radius $R_0$ (see figure \ref{bubblelayer}). 
Then growth of the larger bubble from radius $R_0$ to radius $R$ will
require the given layer to move 
so as to leave a volume equal to the change in the larger bubble's 
volume, which is  $\frac{4\pi}{3}(R^3 - R_0^3)$. So a layer with
initial inner radius $r_0$ must move so that $r_0$ is increased to
$r$, with 
\begin{equation}\label{voldisplaced}
r^3 - r_0^3 = R^3 - R_0^3 
\end{equation}
Hence we obtain a bubble layer's new position $r(r_0)$ as 
\begin{equation}\label{rexact}
r(r_0) = \left( r_0^3 + R^3 - R_0^3 \right)^{1/3}
\end{equation}

In the absence of rearrangements, the linear extension $H$ of bubbles
in a layer with initial
inner radius $r_0$ is 
\begin{equation}
H = \Delta u(r_0+R_B) - \Delta u(r_0) 
\end{equation}
where $\Delta u(r)$ is the distance by which shrunken bubbles at $r$
will move (see figure \ref{bubblelayer}), due to the larger bubbles
growth. So using $\Delta u(r_0) = r(r_0) - r_0$, $\Delta u(r_0
+R_B)=r(r_0+R_B) - (r_0 + R_B)$, and Eq. \ref{rexact}, we expand in
terms of $R_B/r_0$, which to lowest order gives 
\begin{equation}
H \simeq R_B \left( \frac{r_0^2}{r^2} - 1 \right) 
\end{equation}
If we again use Eq. \ref{rexact}, then for small strains or large
distances $r_0$ from a bubble we may expand in terms of $(R^3
-R_0^3)/r_0^3$, to obtain
\begin{equation}\label{8.52}
H \simeq -\frac{2}{3} R_B \left( \frac{R^3 - R_0^3}{r_0^3} \right) + O
\left( \frac{R^3 - R_0^3}{r_0^3} \right)^2   
\end{equation}

Now consider the energy of extending or contracting along one axis
of a single shrunken bubble, with its length changed 
from $R_B$ to $R_B + H$. Restricting ourselves to small strains 
(or equivalently $r_0 \gg R$), we may expand the associated increase
in energy $\Delta E_1(H)$ in terms of a power series in $H$,
\begin{equation}
\Delta E_1 = \sigma \left( \alpha_1 H + \alpha_2 H^2 + O (H^3)
\right)  
\end{equation}
Since we consider an initial state with {\sl unstrained} bubbles, 
equilibrium requires that both positive and 
negative values of $H$ will give positive $\Delta E_1$, 
requiring $\alpha_1=0$. 
So to lowest order $\Delta E_1 = \sigma \left( \alpha_2 H^2 \right)
\sim \sigma H^2$.
  
Initially we have of order $r_0^2dr_0/R_B^3$ shrunken bubbles between
spheres at $r_0$ and $r_0 +dr_0$. So the change in elastic
energy due to a large bubble's growth or shrinking is  
\begin{equation}
\Delta E \sim \int_{R_0}^{\infty} \left
  ( \frac{r_0^2dr_0}{R_B^3} \right) \Delta E_1
\end{equation}
which upon substitution of Eq. \ref{8.52}, $\Delta E_1 = \sigma H^2$,
and integrating gives 
\begin{equation}\label{el2}
\Delta E \sim \sigma \int_{R_0}^{\infty} \frac{r_0^2dr_0}{R_B^3} H^2
\sim \frac{\sigma}{R_B} \frac{(R^3 - R_0^3)^2}{R_0^3} 
\end{equation}
The asymptote of $\Delta E \sim R^6$ for $R \gg R_0$ 
means that in the absence of bubble
rearrangements the elastic energy in the surrounding foam
will ultimately prevent coarsening, and elastically stabilise the system.

\section{Finite Yield Strain}\label{FYS}

The approximation of $\Delta E_1 \sim \sigma H^2$ is strictly only
valid for large $r_0$ or sufficiently small strains. 
For large strains we might expect plastic rearrangements to occur.

Consider the following simple model: the shrunken bubbles may 
elastically support a maximum (yield) strain $y^*$, beyond which
macroscopic rearrangements will occur. Then 
\begin{equation}
y^*\sim H^*/R_B
\end{equation} 
with $H^*$ the plastic threshold for bubble rearrangement. 
In the absence of rearrangements Eq. \ref{8.52} gives  
\begin{equation}\label{8.52.2}
H \sim R_B \frac{|R^3- R_0^3|}{r_0^3}
\end{equation}
so that $H^*$ implicitly defines a radius $r_0^*$ within which
rearrangements occur, but beyond which the medium behaves
elastically. 
So from Eq. \ref{8.52.2}, we may define $r_0^* \equiv
({R_B}/{H^*})\left(|R^3-R_0^3| \right)$, which  since $H^* \sim y^*
R_B$ gives  
\begin{equation}\label{r0*1}
{r_0^*}^3 \sim \frac{|R^3 -R_0^3|}{y^*} 
\end{equation}
For a plastic region to exist around a
bubble, we require $r_0^* > R$; Eq. \ref{r0*1} then requires
$\left| R^3 - R_0^3 \right| > y^* R^3$. 

Whether sufficient growth is possible for this threshold to be reached depends
on the radius $R'$ at which
bubble growth would halt, in a purely elastic system as considered in the previous section.  If the inequality $\left| R'^3 - R_0^3 \right| > y^* R'^3$ is not satisfied (as occurs for large enough $y^*$) 
one requires  a large fluctuation, either thermally or in the initial
condition, to establish a plastic zone around a droplet, allowing it to coarsen further. In that case the foam is at least metastable and in practice
this may be sufficient.

To get a stability condition, we consider next the elastic energy stored
around a droplet that has grown from size $R_0$ to size $R$.
Taking deformations within any plastic region to be of order $H^*$,
we may calculate the elastic energy as 
\begin{equation}
\Delta E \sim \sigma \int_{R_0}^{r_0^*} \frac{r_0^2dr_0}{R_B^3}{H^*}^2 +
\sigma \int_{r_0^*}^{\infty} \frac{r_0^2dr_0}{R_B^3} H(r_0)^2 
\end{equation}
giving 
\begin{equation}\label{elastopE}
\Delta E \sim y^* \frac{\sigma}{R_B} \left( \left| R^3 - R_0^3
  \right| - y^*R_0^3 \right) + y^* \frac{\sigma}{R_B} \left| R^3 -
  R_0^3 \right|
\end{equation} 
where, since $y^* < \left| R^3 - R_0^3 \right|/R_0^3$, both terms 
on the right are positive. 

We now assume that a grown bubble adopts a state of mechanical
equilibrium with the surrounding shrunken-bubble sea, on a timescale
fast compared to any coarsening process.
Then for a reversible fluctuation in bubble volume by $dV$,
the work done by the bubble's gas equals the
work done on the bubbles surrounding environment. Hence 
\begin{equation}\label{elP}
P_idV \sim \left( P + \Pi + \frac{\partial(\sigma A)}{\partial
  V} 
+ \frac{\partial(\Delta E)}{\partial V}  \right) dV 
\end{equation}
So differentiating Eq. \ref{elP}, with $\Delta E$ given by
Eq. \ref{el2} for  $\left| R^3 - R_0^3 \right| < y^* R^3$ and by
Eq. \ref{elastopE} for  $\left| R^3 - R_0^3 \right| > y^* R^3$, we get 
\begin{equation}
P_i \sim P + \Pi + \frac{\sigma}{R_i} + 
\left[
\begin{array}{ll}
\frac{\sigma}{R_B} \frac{R^3 - R_0^3}{R_0^3} &\;\;\;\;\;\left| R^3 - R_0^3
  \right| < y^* R^3 
  \\  y^*\frac{\sigma}{R_B} \frac{R^3 - R_0^3}{|R^3 - R_0^3|}
  & \;\;\;\;\;\left| R^3 - R_0^3 \right| > y^* R^3 
\end{array}\right.
\end{equation}
This means that any bubble that has successfully coarsened to become a bulk gas phase (whose pressure clearly cannot be infinite as $R \to \infty$) has 
overcome the plastic threshold and has a pressure $P_b$ obeying
\begin{equation}\label{Pbplastic}
P_b -P-\Pi \simeq y^*\frac{\sigma}{R_B}
\end{equation}

Now we are able to estimate a stability threshold for a foam that combines a finite amount of trapped species with a finite yield strain (under the simplified initial condition we chose above, that the initial state has a dilute array of partially grown bubbles in an unstrained, shrunken-bubble sea).
In this case the lowest possible free energy obtainable by coarsening is that of a
bulk gas at pressure $P_b$. 
So unless the initial pressure $P_i$ of grown bubbles 
exceeds $P_b$, then the foam {\sl must} be stable\cite{ff2}. 
An initially unstrained bubble has 
\begin{equation}
P_i -P -\Pi \simeq \frac{\sigma}{R_0} 
\end{equation}
so instability requires $P_i > P_b$, and hence 
\begin{equation}
\frac{\sigma}{R_0} \grtapprox \frac{y^*\sigma}{R_B} 
\end{equation}
So since $R_B \ll R_0$, such a foam may only be unstable if $y^* \ll 1$. 
For example, considering foams formed by the osmotic compression of a
dilute partially coarsened foam, then if $R_0$ is small enough that
$R_0<R_B/y^*$ then coarsening may occur, but if it has already
coarsened sufficiently that $R_0>R_B/y^*$, then after osmotic
compression it will be elastically stabilised. 

\section{Conclusions}\label{conclusions}

We have considered dilute foams with spherical bubbles, and nondilute
foams in which bubbles are compressed by an osmotic pressure $\Pi$
(which distorts bubbles and increases their gas pressure). 
Prior to osmotic compression, variations in bubble pressures are of
order $\sigma/R_i$, so bubbles may only support large osmotic pressures ($\Pi \gg \sigma/R_i$)
by deforming and pressing on one another. 
Pressure differences between such osmotically compressed bubbles generally remain
limited to $\sigma/R_i$ (with a possible exception in the case of extreme
polydispersity in initial droplet size). 
If coarsening occurs, a coexisting bulk gas is created that has 
pressure $P_b=P+\Pi$, with $P$ the ambient atmospheric pressure. 
We defined the excess in a bubble's pressure above that of such a bulk gas by $P_i^G \equiv
P_i-P-\Pi$, and argued that $P_i^G \sim \sigma/R_i$. 
This was confirmed quantitatively and explicitly for a monodisperse 2D
model of foam in Section \ref{2Dmodel}. $P_i^G$, which we call the geometric
pressure, provides the driving force behind coarsening in the case of a compressed foam or emulsion; in general it is very different
from the Laplace pressure but reduces to it in the dilute limit.
For dilute 2D and 3D systems and also for a 
concentrated but monodisperse 2D foam, $P_i^G$ can be
found explicitly.
This allows one to find  rigorous conditions (Eqs. \ref{dilcond}, \ref{2Ddilcond}, and 
\ref{NsB2D}) for their stability. 
More generally we have argued that for a reasonably large $\Pi$ and a
reasonably narrow distribution of $N_i^T$, osmotic stabilisation should
generally be possible, and typically requires $P_i^T \grtapprox
\sigma/R_i$.  

The morphology of an insufficiently stabilised foam (containing inadequate
levels of a trapped species), is determined by the `excess
volume fraction' of dispersed phase; this will form a
bulk gas phase if the system is allowed to reach full equilibrium ({\sl i.e.,} subject only to the constraint that the trapped species stay in their designated droplets and coalescence is absent, which with trapped species
present will ensure conservation of droplet number). 
The kinetics of coarsening will be determined either by the rate of
diffusion of gas between bubbles or the rate at which shrunken
bubbles rearrange, with the slowest process determining the bubble 
growth rate. 
We studied coarsening of a small excess volume fraction, finding
diffusion limited growth to result in $\bar{R} \sim t^{1/3}$ (as
opposed to $\bar{R}\sim t^{1/2}$ as arises in foams without trapped
molecules\cite{Glazier}),
and viscous limited growth to give $\bar{R} \sim t$. 
For typical parameters we expect diffusion limited growth to be more common, with viscous
limited growth more likely to be observed if the liquid 
viscosity is high, the temperature is high, and the pressure is
low\cite{ff3}.  
Under conditions of negligible bubble rearrangements, the buildup of elastic
stresses among the sea of shrunken, stabilised bubbles will arrest coarsening of well-separated large bubbles and `elastically stabilise' the entire foam. 
This can sometimes occur when there is a finite yield strain;
whether a foam will be elastically stabilised is then
determined by the foam's initial state (and the yield strain $y^*$). 
Thus osmotic compression of a partially coarsened
{\sl dilute} foam, may {\sl halt} coarsening and elastically stabilise
the foam; the further the foam has coarsened initially, the more
likely this is to occur.

Our work applies, substantially unchanged, to concentrated emulsions. 
As with nondilute foams the disjoining energy between repelling
droplets in emulsions may usually be neglected \cite{BuzzaCates},
and the osmotic pressure of trapped molecules in droplets
equals the partial pressure of trapped gas
molecules in bubbles (so long as both are treated as ideal \cite{Webster,WebsterTh}). 
Hence if we replace the volume per gas molecule $v_g=\tilde{P}/kT$
with the molecular volume of disperse phase in droplets $v_b$, and
also replace $N_i^{sB}=(kT/\tilde{P})V_i^B$ with
$N_i^{sB}=(1/v_b)V_i^B$, then most of 
the above work applies equally well to
concentrated emulsions.


\appendix

\section{Force Balance}\label{Fbalance}

We adopt a similar approach to one previously used by
Princen\cite{Princen1,Princen2}, which will enable us to determine the   
order of magnitude of $P^G_i$ when bubble sizes are polydisperse. 
Unlike the one in the main text, this approach is equally 
valid for compressible and incompressible bubbles.

\begin{figure}[htd]
\begin{center}
\leavevmode
\epsfig{file=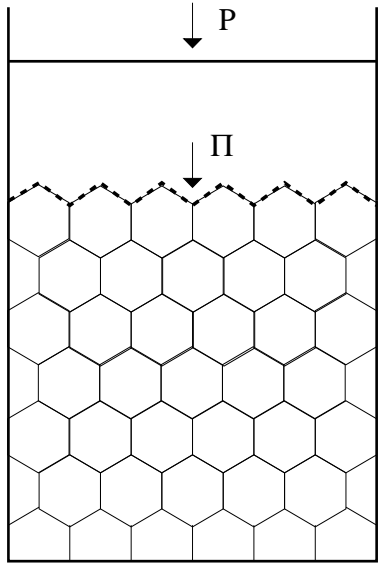,width=5cm,height=7.5cm}    
\end{center}
\caption{Imagine a flexible semipermeable membrane initially between
  bubble layers, which subsequently hardens and enables you to remove
  bubbles above the membrane, while maintaining a pressure
  $\Pi$ on the 
  bubbles below. At equilibrium
  the forces above and
  below the membrane must balance. The same thought experiment may
  also be applied to polydisperse bubbles.}\label{monobrane}   
\end{figure}

Consider a semipermeable membrane with a shape which follows the top
surface of a line of hexagonal bubbles (see figure \ref{monobrane}),
so that bubbles remain hexagonal in shape. Then since the membrane
doesn't move, the total force on the membrane due to $\Pi + P$ must
balance the total force on the membrane acting from below. Then we
consider a typical bubble adjacent to the membrane (see figure
\ref{monobrane}).
Such a bubble will have a geometry as in figure \ref{nhexpict} but
with the bubble rotated by $\pi/6$ to leave two flat edges vertical. 
The force acting from below the membrane is $P_i$
multiplied  by the
projection onto the horizontal axis of the flat bubble
edges($P_i\times 2(l\sqrt{3}/2)$), plus $P$
multiplied by the projection of the Plateau borders onto the
horizontal axis ($P\times 2r$). 
The force acting from above the membrane is simply $P+\Pi$ multiplied
by the bubble's width $2(l\sqrt{3}/2) + 2r$. 
Then for the forces above and below to balance we require\cite{f7} 
\begin{equation}
P_i2\frac{l\sqrt{3}}{2} + P2r = \left( \Pi + P \right) \left
  (  2\frac{l\sqrt{3}}{2} + 2r \right) 
\end{equation}
Writing $P_i = P +\sigma/r$ and rearranging, we obtain
\begin{equation}
\Pi = \frac{\sigma}{r} - \frac{6\sigma}{3\sqrt{3}l + 6r}
\end{equation}
which writing in terms of $\mathcal{A}$ and $\mathcal{A}_l$ and comparing with
Eq. \ref{PG2} gives $\Pi = {\sigma}/{r} - P^G_i$ 
as before. Note that since the method may be applied to compressible
systems, both $P_i$ and $P^G_i$ for the monodisperse 2D model will
remain the same for both compressible and incompressible bubbles. 

We now generalise the argument to a polydisperse system. The
projections of the bubble faces and plateau borders onto the
horizontal axis are written as $l_i^{P_i}$ and $l_i^P$
respectively. Then considering a suitably shaped membrane, force
balance requires
\begin{equation}
\langle P_i l_i^{P_i} \rangle_i + P \langle l_i^P \rangle_i =
(\Pi+P)\left( \langle l_i^{P_i} \rangle_i + \langle l_i^P \rangle_i
\right) 
\end{equation}
So writing $P_i = P + \sigma/r_i$ and restricting ourselves to fairly
dry foams with $r_i \simeq r_j$, then $\langle l_i^{P_i}/r_i \rangle_i
\simeq \langle l_i^{P_i} \rangle_i/r$ where $r\equiv \langle r_i
\rangle_i$, and we can obtain  
\begin{equation}
\Pi \simeq \frac{\sigma}{r} - \frac{\sigma}{r} \frac{\langle l_i^P
  \rangle_i}{\langle l_i^{P_i} \rangle_i+\langle l_i^P \rangle_i} 
\end{equation}
Now we note that since $l_i^P\sim r$ and $l_i^{P_i}\sim \bar{R}$, where
$\bar{R}\equiv \langle \sqrt{\bar{\mathcal{A}}}/\pi \rangle_i$, then 
$\Pi \sim {\sigma}/{r} - {\sigma}/{\bar{R}}$ 
so that if $P^G_i \equiv P_i - P - \Pi$ then $\langle P^G_i \rangle
\sim \sigma/\bar{R}$.  

In 3D the same argument but with projected areas $\langle A_{i}^P
\rangle \sim \bar{R}\times r$, and $\langle A_{i}^{P_i} \rangle \sim 
\bar{R}^2$ leads to $\Pi \sim {\sigma}/{r} -
{\sigma}/{\bar{R}}$ which for $P^G_i = P_i -P - \Pi$ gives 
\begin{equation}\label{avPG}
\langle P^G_i \rangle \sim \sigma/\bar{R} 
\end{equation}
Finally we note that in higher dimensions the equivalent argument will
continue to give $\langle P^G_i \rangle \sim \sigma/\bar{R}$.

\section{Geometry of Monodisperse, 2D Bubbles}\label{bubblegeom}

The area of a hexagon of side $l$ is six times the area of the equilateral
triangles of which it is composed. The area of each triangle is
$(l/2)h$, with $h=l\sin (\pi/3)=l\sqrt{3}/2$. Hence we obtain the
area of a hexagon as $\mathcal{A}=({3\sqrt{3}}/{2}) l^2$. 
Its interfacial length is simply given by $\mathcal{L}=6l$. 

\begin{figure}[htd]
\begin{center}
\epsfysize=48mm
\epsfxsize=51mm
\leavevmode
\epsffile{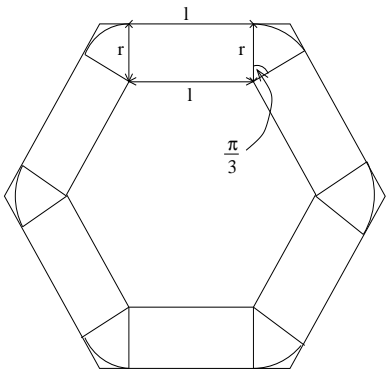}
\end{center}
\caption{Nearly hexagonal foam bubble with flat edges of length l, and
  radii of curvature $r$. The perpendicular distance between two flat
  edges is $l\sqrt{3}+2r$.}\label{nhexpict} 
\end{figure}

We now describe a nearly hexagonal bubble in terms of the length of the
flattened faces $l$, and the radius of curvature of the plateau border
$r$ (see figure \ref{nhexpict}). 
Then the area $\mathcal{A}$ of a nearly hexagonal bubble is given by
the sum of six 
sections of a circle, each of angle $\pi/3$ and radius $r$, plus six
rectangles of length $l$ and height $r$, plus the area of a hexagon of
side length $l$. So 
\begin{equation}\label{area_nhex}
\mathcal{A} = \pi r^2 +6lr + \frac{3\sqrt{3}}{2} l^2 
\end{equation}
and the surface length obeys 
\begin{equation}\label{length_nhex}
\mathcal{L} = 6l + 2\pi r
\end{equation}

\section{Bubbles in Plateau Borders}\label{polydappendix}

What is required for states such as that in figure
\ref{voldependence} to exist? 
Using $P_{i}^s=P + 2\sigma/R_{i} - P_i^T$ and the ideal gas law (for
the trapped gas), the pressure of soluble gas in a spherical bubble
residing in a Plateau border is $P_{i}^s = P +  
2\sigma/R_{i} -  N_{i}^TkT/(4\pi/3)R_{i}^3$, which has a maximum
value of $P+4(8\pi)^{1/2}\sigma^{3/2}/9(kTN_{i}^T)^{1/2}$. 
So in a system which coarsens with $\bar{P}^s\rightarrow P+ \Pi$, the
existence of spherical bubbles in Plateau borders requires 
\begin{equation}
P+\frac{4}{9} \sqrt{\frac{8\pi\sigma^3}{kTN_{i}^T}} \geq P + \Pi 
\end{equation}
Hence such bubbles will need a sufficiently small quantity of trapped
species that
\begin{equation}
N^T_{i} \leq \left( \frac{8}{9} \right)^2 \frac{2\sigma^3}{kT}
\frac{1}{\Pi^2} 
\end{equation}
which writing $\Pi$ in terms of the Laplace pressure of a bulk gas,  
$\Pi = 2\sigma/r_b$, requires 
\begin{equation}
N^T_{i} \leq \left( \frac{8}{9} \right)^2 \frac{\sigma r_b^2}{2kT} \sim
\frac{\sigma r_b^2}{kT} 
\end{equation}
We may compare $N_{i}^T$ with that for a typical stable (and
compressed) bubble of radius $\bar{R}$, containing $\bar{N}^T$ trapped
molecules.  
Then since $P^T(\bar{N}^T,\bar{R})=\bar{N}^TkT/(4\pi/3)\bar{R}^3 \sim
\sigma/\bar{R}$ one requires  
\begin{equation}
\bar{N}^T \sim \frac{\sigma \bar{R}^2}{kT} 
\end{equation}
A spherical bubble in a Plateau border has 
\begin{equation}
N^T_{i} \lessapprox \bar{N}^T \frac{r_b^2}{\bar{R}^2} 
\end{equation}
Considering $\Pi$ for which $r_b/\bar{R} \sim 10^{-1}$ or
smaller, then we require $N_{i}^T \lessapprox 10^{-2}\bar{N}^T$. 
So for a reasonable $\Pi$ of order $\sigma/r_b$, and a reasonably narrow
distribution for $N_i^T$, then there will be a negligible volume
fraction of bubbles residing in Plateau borders.

\section{Stability Requirements}\label{ReqApp}

We examine the form of
$\bar{N}_i^{sB}(t)$, the number of soluble species per bubble below
which a stable foam is ensured.
Noting that $\bar{N}_i^{sB}(t)= \langle
N_i^{sB}(t) \rangle_i$, we write it out in full to obtain,
\begin{equation}\label{90}
\langle N_i^{sB} (t) \rangle_{i} = \frac{\tilde{P}}{kT}
\left( \frac{kT}{\sigma} \right)^{\mathcal{D}/(\mathcal{D}-1)}
\left< \left( \frac{N_i^T}{\Gamma_i} \right)^{\mathcal{D}/(\mathcal{D}-1)}
\right>_{i}
\end{equation}
We note that $\mathcal{D} \geq 2$ so that 
\begin{equation}\label{91}
\left< \left( \frac{1}{\Gamma_i} \right)^{\mathcal{D}/(\mathcal{D}-1)}
\right>_i \geq
\frac{1}{\langle \Gamma_i \rangle_i^{\mathcal{D}/(\mathcal{D}-1)} }
\end{equation}
We assume that $\Gamma_i$ and $N_i^T$ are uncorrelated, that is for a
given bubble volume we assume that variations in $P_i^G$ are
independent of $N_i^T$. Then using
Eqs. \ref{90} and \ref{91}, we have
\begin{equation}\label{NsBbound}
\langle N_i^{sB} (t) \rangle_{i} \geq \frac{\tilde{P}}{kT}
\left( \frac{kT}{\sigma} \right)^{\mathcal{D}/(\mathcal{D}-1)}
\frac{\langle {N_i^T}^{\mathcal{D}/(\mathcal{D}-1)}
  \rangle_{i}}{\langle \Gamma_i
\rangle_i^{\mathcal{D}/(\mathcal{D}-1)}}
\end{equation}

In Section \ref{Fbalance} we found that $\langle P^G \rangle \sim
\sigma/\bar{R}$, which suggests $\bar{\Gamma}$ is of order a constant
$\bar{\Gamma}_C$. Then since 
\begin{equation}\label{Ansatzbound}
\langle N_i^{sB} (t) \rangle_{i} \geq \frac{\tilde{P}}{kT}
\left( \frac{kT}{\sigma} \right)^{\mathcal{D}/(\mathcal{D}-1)}
\frac{\langle {N_i^T}^{\mathcal{D}/(\mathcal{D}-1)}
  \rangle_{i}}{\bar{\Gamma}_C^{\mathcal{D}/(\mathcal{D}-1)}} 
\end{equation}
a foam formed with an initial number of soluble species
$\bar{N}^{s0}$ such that 
\begin{equation}\label{NsBbound1}
\frac{\tilde{P}}{kT}
\left( \frac{kT}{\sigma} \right)^{\mathcal{D}/(\mathcal{D}-1)}
\frac{\langle {N_i^T}^{\mathcal{D}/(\mathcal{D}-1)}
  \rangle_{i}}{\bar{\Gamma}_C^{\mathcal{D}/(\mathcal{D}-1)}}  
\geq \bar{N}^{s0} 
\end{equation}
will be stable. Similarly if $\bar{\Gamma}$ is bounded by
$\bar{\Gamma}_{max}$, then a stability condition is obtained by
replacing $\bar{\Gamma}_C$ with $\bar{\Gamma}_{max}$ in
Eq. \ref{NsBbound1}. 
Alternately, if $\bar{\Gamma}$ is a decreasing function of time, then
$\bar{\Gamma}(0) \geq \bar{\Gamma}(t)$ and hence
$1/\bar{\Gamma}(t)^{\mathcal{D}/(\mathcal{D}-1)} \geq
1/\bar{\Gamma}(0)^{\mathcal{D}/(\mathcal{D}-1)}$, so 
provided 
\begin{equation}\label{NsBbound3}
\frac{\tilde{P}}{kT}
\left( \frac{kT}{\sigma} \right)^{\mathcal{D}/(\mathcal{D}-1)}
\frac{\langle {N_i^T}^{\mathcal{D}/(\mathcal{D}-1)}
  \rangle_{i}}{\bar{\Gamma}(0)^{\mathcal{D}/(\mathcal{D}-1)}}  
\geq \bar{N}^{s0} 
\end{equation}
then we could again guarantee a stable foam. 
In contrast to the above cases, if $\bar{\Gamma}$ {\sl increases}
without bound, then $\bar{N}^{sB}(t)$ will {\sl decrease}
without bound, and a stability condition will never strictly exist. 
The only way a stability condition can fail to exist
is if $\bar{\Gamma}$ can increase in time {\sl without bound}.

\section{Table of Notation}\label{notation}

${\cal A}_i, {\cal A}_{li}, {\cal A}_i^B, {\cal A}_{Hi}^B$: area of two-dimensional bubble $i$; area of surrounding liquid (within hexagonal unit cell); area at which such a bubble (containing trapped species) coexists osmotically with a bulk gas phase; the same, but for a completely dry hexagonal foam.

$C(R)$: equilibrium concentration of dissolved gas adjacent to a bubble of radius $R$.

$c(r)$: mean concentration of dissolved gas within the continuous phase at distance $r$ from the
centre of a bubble.   

$D$: diffusivity of dissolved gas within the continuous phase.

${\cal D}$: dimensionality of space. 

$d$: thickness of films of continuous phase separating faces of adjacent bubbles.

$H, H^*$: local extensional or compressive strain of bubbles within a sea of shrunken bubbles; its maximum value before yielding occurs.

${\bf J}_V$: volume flux of soluble gas around coarsening bubble.

$K$: flux/pressure coefficient for gas permeation across a fluid film.

${\cal L}_i$: the perimeter length of bubble $i$ in a two-dimensional foam.

$l$: length of a flat interface between adjacent bubbles in a two-dimensional hexagonal monodisperse foam.

$N_i^T, N_i^s, N_i^{sB}$: number of trapped molecules in $i$th bubble; number of soluble molecules within $i$th bubble; the number of soluble molecules within $i$th bubble when coexisting with a bulk gas phase.

$\bar N^{s0}, \bar N^{sB}, \bar N^{sC}$: average number of soluble gas molecules per bubble; average number of soluble molecules per bubble within the stable (noncoarsening) population; average number of soluble molecules per bubble within the (unstable) coarsening population.

$n_b^0, n_b^S, n_b^C$: total number density of bubbles in the system; number density of shrunken bubbles; number density of coarsening
bubbles.

$P$: ambient pressure at which a foam is maintained.

$\tilde P$: defined as $\tilde P\equiv P+\Pi$.

$P_b^s$: partial pressure of soluble gas in a bulk gas phase created by a coarsening process, and equal to the total gas pressure in this phase.

$P_i, P_i^s,P_i^T$: gas pressure within $i$th bubble; partial pressure of soluble species within it; partial pressure of trapped species within it.

$P_i^G$: geometric pressure of $i$th bubble (see text for definition of geometric pressure).

$P_s(r,t)$: mean partial pressure of soluble gas in shrunken bubbles at radius $r$ from a growing bubble.

$R_i, \bar R, R_0, R(t), R', R_B$:  radius of $i$th bubble; mean bubble radius; initial radius of a coarsening bubble; its radius at time $t$; radius at which its growth is arrested in absence of rearrangement; radius of a shrunken bubble at or near coexistence with a bulk gas phase.

$R_G$: the gas constant.

$r, r_0, r_0^*$: radial distance of a given layer of shrunken bubbles from the centre of a growing bubble; its initial value; the latter quantity for bubbles whose local strain coincides with the yield strain of the shrunken bubble sea.  

$r_i$: radius of curvature at Plateau border of $i$th bubble.

$\dot S_D, \dot S_V$: entropy production due to dissipative and viscous fluxes during coarsening.

$u(r),v(r)$: radial velocity of dissolved gas molecules at distance $r$ from the centre of a coarsening bubble; radial velocity of shrunken-bubble fluid.

$V_i, V_i^B$: volume of bubble $i$; its volume at osmotic coexistence with a bulk gas phase.

$v_g,v_{Mg}$: molecular volume of disperse phase species; its molar volume. 

$y^*$: yield strain of the shrunken bubble sea.

$\alpha_1, \alpha_2$: numerical coefficients (dimensionless).

$\Delta E, \Delta E_1(H)$: change in elastic energy of shrunken bubble sea on growth of a coarsening bubble; change in energy of a single shrunken bubble due to deformation $H$.

$\Delta \mu_i$: chemical potential difference, $\mu_i^s - \mu_i^B$.

$\epsilon$: dimensionless supersaturation $(P^s(\infty,t)-\tilde P)/\tilde P$.

$\eta$: viscosity of continuous phase.

$\phi,\phi_0$: volume fraction of bubbles; its value at first contact.

$\Gamma_i, \bar \Gamma$: for the $i$th bubble, the value of $P_i^G V_i^{1/{\cal D}} \sigma$; its mean value. 

$\dot \gamma$: shear rate.

$\mu_i^s,\mu_b^s$: chemical potential of soluble disperse phase in $i$th bubble;
chemical potential of a bulk phase of the same species (at pressure $P$).

$\Pi$: osmotic pressure in a foam.

$\sigma$: surface tension of films in a foam.



\begin{thebibliography}{99}
{\renewcommand{\baselinestretch}{2.0}\small\normalsize

\bibitem{Higuchi} Higuchi, W.I. ;  Misra, J.
  {\sl J. Pharm. Sci.} {\bf 1962} {\sl 51}, 459.

\bibitem{Hallworth} Hallworth, G.W. ; Carless, J. In
  {\sl Theory and Practice of Emulsion Technology} {\bf 1976}
  (A.L. Smith, Ed.), Academic Press, London/New York. p305.

\bibitem{Davis} Davis, S.S. ; Smith, A. In 
  {\sl Theory and Practice of Emulsion Technology} {\bf 1976}
  (A.L. Smith, Ed.), Academic Press, London/New York. p325.

\bibitem{Davis2} Davis, S.S. ; Round, H.P. ; Purewal, T.S. 
  {\sl J. Coll. Int. Sci.} {\bf 1980} {\sl 80}, No. 2, 508. 

\bibitem{Kab87a} Kabalnov, A.S. ; Pertsov, A.V. ; Shchukin, E.D.
  {\sl J. Coll. Int. Sci.} {\bf 1987} {\sl 118}, No. 2, 590. 

\bibitem{Kab87b} Kabalnov, A.S. ; Pertsov, A.V. ; Shchukin, E.D. 
  {\sl Colloids. Surf.} {\bf 1987} {\sl 24}, 19--32. 

\bibitem{Reiss} Reiss, H. ;  Koper, J.M. 
  {\sl J. Phys. Chem.} {\bf 1995} {\sl 99}, 7837--7844.

\bibitem{Kuehmann} Kuehmann, C.J. ; Voorhees, P.W. {\sl
             Metall. Trans. A } {\bf 1996} {\sl 27A}, 937--943. 

\bibitem{Webster} Webster, A.J. ; Cates, M.E. {\sl Langmuir} {\bf
    1998} {\sl 14}, 2068--2079. 

\bibitem{Ostwald}  Ostwald, W. {\sl Z. Phys. Chem.} (leipzig)
            {\bf 1900} {\sl 34}, 295.



\bibitem{Lifshitz} Lifshitz, I.M. ; Slyozov, V.V. 
            {\sl Phys. Chem. Solids.} {\bf 1961} {\sl 19} 35--50.

\bibitem{Wagner} Wagner, C. {\sl Z. Elektrochemie} {\bf 1961} {\sl
    65}, 581--591. 

\bibitem{WebsterTh} Webster, A.J. {\sl Coarsening and Osmotic
    Stabilisation of Emulsions and Foams} {\bf 2000} Ph.D. Thesis,
    Edinburgh University, Scotland. 



 
\bibitem{Falls} Falls, A. H.; Lawson, J. B.; Hirasaki, G. J. {\sl J.
Pet. Technol.} {\bf 1988} p95 

\bibitem{Gandolfo} Gandolfo, F. G. ; Rosano, H. L. {\sl J. Coll. Int.
Sci.} {\bf 1997} {\sl 194} 31--36.

\bibitem{Parr} Parr, D. {\sl Physics World} {\bf 1989} {\sl September},
p21

\bibitem{Bamforth} Bamforth, C. W. In {\sl Food Colloids} {\bf 1989} ed.
Bee, R.D.; Richmond, P. ; Mingins, J. (London: Royal Society of
Chemistry) p48.

\bibitem{Weaire_Pageron} Weaire, D. ; Pageron, V. {\sl Philos. Mag.
Lett.} {\bf 1990} {\sl 62}, No. 2, 417--421.

\bibitem{ultrasound} Kabalnov, A.; Klein, D.; Pelura, T.; Schutt, E.; Weers, J.;
{\sl Ultrasound in Med. Biol.} {\bf 1998}, {\sl 24} 739--749. 

\bibitem{Stone} Hilgenfeldt, S.; Koehler, S. A.; Stone, H. A. {\sl The dynamics
of coarsening foams: accelerated and self-limited drainage}, preprint, 2000.




\bibitem{Princen1} Princen, H.M. {\sl J. Coll. Int. Sci.} {\bf 1979}
  {\sl 71}, No. 1, p55--66. 

\bibitem{Princen2} Princen, H.M. {\sl Langmuir} {\bf 1986} 
  {\sl 2}, No. 4, p519--524.  

\bibitem{Princen3} Princen, H.M. {\sl Langmuir} {\bf 1988} 
  {\sl 4}, No. 1, p164--169. 

\bibitem{Rowlinson} Rowlinson, J.S. ; Widom, B. In {\sl Molecular Theory
  of Capillarity} {\bf 1989} Oxford University Press, Walton
  Street, Oxford. 

\bibitem{f1} Nondilute foams resemble concentrated emulsions,  
  which consist of incompressible emulsion droplets which
  are pressed together by the action of a semipermeable membrane
  (through which only continuous phase may pass).


\bibitem{BuzzaCates} Buzza, D.M.A.; Cates, M.E. {\sl Langmuir} {\bf
1993} {\sl 9}, 2264--2269. 

\bibitem{f2} Since $P_b^T$ is negligible for a bulk gas formed via
  coarsening, we take $P_b^T=0$. 

\bibitem{f3}Assumed to be much more rapid than diffusion of gas
  between bubbles.


\bibitem{Radjai} Radjai, F. ; Wolf, D.E. ; Jean, M. ; Moreau, J--J. 
  {\sl Phys. Rev. Lett.} {\bf 1998} {\sl 80}, No. 1, 61--64.  

\bibitem{Fragile} Cates, M.E. ; Wittmer, J.P. ; Bouchaud, J--P. ;
  Claudin, P. {\sl Phys. Rev. Lett.} {\bf 1998} {\sl 81}, No. 9,
  1841--1844. 


\bibitem{f4} The present argument ignores highly polydisperse systems
  in which very small bubbles may be uncompressed  ($\Pi=0$), by
  residing in bubble's Plateau borders. 


\bibitem{Durian91} Durian, D.J.; Weitz, D.A.; Pine, D.J. {\sl
    Phys. Rev. A} {\bf 1991} {\sl 44}, No 12, R7902--R7905.

\bibitem{Durian97} Durian, D.J.; {\sl Current Opinion in Colloid \&
    Interface Science} {\bf 1997} {\sl 2}, 615--621. 








\bibitem{Bray} Bray, A.J. {\sl J. Adv. Phys.} {\bf 1994}, {\sl 43},
  357--459. 

\bibitem{Glazier} Glazier, J.A. ; Weaire, D. {\sl J. Phys. Condens.
Matter} {\bf 1992} {\sl 4} p1867--1894.


\bibitem{Reif} Reif, F. In {\sl Fundamentals of Statistical and Thermal
    Physics} {\bf 1985} International Edition, McGraw--Hill
    Co--Singapore, p567. 


\bibitem{vonfootnote} von Neumann's law gives in two dimensions the rate of change of area of the $i$th bubble as
$d{\cal A}_i/dt = K\sigma\pi(n_i-6)/3$ where $n_i$ is its number of edges.
With trapped species present, one has $d{\cal A}_i/dt = K\sigma\pi(n_i-6)/3 +
K\sum_j(P_i^T-P_j^T)l_{ij}$ where $j$ is a neighbouring cell and $l_{ij}$
the length of the common edge. The sign of the extra term depends on the
volumes of the neighbouring cells (and the number of trapped species they contain) so that static equilibrium may be reached in a system where not all
cells have $6$ neighbours. See \cite{WebsterTh}.


\bibitem{CRC} {\sl CRC Handbook of Chemistry and Physics}, 74th edition,
  {\bf 1993} CRC press, inc., Ed. DL Lide.

\bibitem{ff1} We note that molecules in dilute {\sl emulsion
    droplets} have a much smaller value of $v_{Mg}$ than do molecules
  in gaseous foam--bubbles. Hence even if there is a significant
  discrepancy between the molecular volumes of dissolved disperse phase
  and disperse phase in droplets, coarsening of dilute emulsions will
  be diffusion limited.


\bibitem{ff2} We continue
to take the pressure of trapped gas in grown bubbles as negligible, so
that a grown bubble's pressure is its soluble gas pressure.


\bibitem{ff3} It is interesting that the viscous limited and diffusion
  limited growth laws Eqs. \ref{mce10.22}, \ref{mce10.30} may be
  predicted (up to constant prefactors), by the most naive order of
  magnitude arguments for the phase ordering kinetics of binary fluids
  (see \cite{Bray}, page $375$). However such arguments {\sl do not}
  correctly predict when each regime will apply, this depends on
  the structure of the shrunken--bubble fluid.  



\bibitem{f7} Where the force acting is pressure multiplied by distance
  over which it acts, projected onto the horizontal axis.



}

\end{thebibliography}
\end{document}